\documentclass[10pt, conference, compsocconf]{IEEEtran}

\usepackage{amsmath}
\usepackage{graphicx}
\usepackage{listings}
\usepackage{caption}
\usepackage{epstopdf}
\usepackage{url}
\usepackage[justification=centering,font={footnotesize}]{caption}
\usepackage{ftnxtra}
\usepackage{threeparttable}
\usepackage{balance}
\usepackage{pstricks}
\usepackage{pst-rel-points}
\usepackage{pst-node}
\usepackage{courier}

\makeatletter
\def\@copyrightspace{\relax}
\makeatother

\newcounter{lcount}
\newenvironment{myenumerate}{
\begin{list}{\arabic{lcount}.}
{\usecounter{lcount}\setlength{\topsep}{1mm}\setlength{\itemsep}{0.25mm}
\setlength{\parsep}{0.1mm}
\setlength{\itemindent}{0mm}\setlength{\partopsep}{0mm}
\setlength{\labelwidth}{15mm}
\setlength{\leftmargin}{4mm}}}{\end{list}}

\newcommand{\cmt}[1]{}

\begin{document}

\newcommand{\blind}[1]{}

\title{
 Improving GPU Performance Through Resource Sharing \vskip -6mm}%
\author{%
\IEEEauthorblockN{Vishwesh Jatala}
\IEEEauthorblockA{%
Department of CSE, IIT Kanpur\\
vjatala@cse.iitk.ac.in}%
\and
\IEEEauthorblockN{Jayvant Anantpur}
\IEEEauthorblockA{%
SERC, IISc, Bangalore \\
jayvant@hpc.serc.iisc.ernet.in}%
\and
\IEEEauthorblockN{Amey Karkare}
\IEEEauthorblockA{%
Department of CSE, IIT Kanpur\\
karkare@cse.iitk.ac.in}%
}
\maketitle
\begin{abstract}

Graphics Processing Units (GPUs) consisting of Streaming Multiprocessors (SMs) achieve high throughput by running a large number of threads and context switching among them to hide execution latencies.  The number of thread blocks, and hence the number of threads that can be launched on an SM, depends on the resource usage--e.g. number of registers, amount of shared memory--of the thread blocks. Since the allocation of threads to an SM is at the thread block granularity, some of the resources may not be used up completely and hence will be wasted.

We propose an approach that shares the resources of SM to utilize the wasted resources by launching more thread blocks. We show the effectiveness of our approach for two resources: register sharing, and scratchpad (shared memory) sharing. We further propose optimizations to hide long execution latencies, thus reducing the number of stall cycles. We implemented our approach in GPGPU-Sim simulator and experimentally validated it on several applications from 4 different benchmark suites: GPGPU-Sim, Rodinia, CUDA-SDK, and Parboil. We observed that with register sharing, applications show maximum improvement of 24\%, and average improvement of 11\%. With scratchpad sharing, we observed a maximum improvement of 30\% and an average improvement of 12.5\%.

\end{abstract}

\begin{IEEEkeywords}
Register Sharing; Scratchpad Sharing; Warp Scheduling; Thread Level Parallelism
\end{IEEEkeywords}

\IEEEpeerreviewmaketitle

\vspace*{-1mm}
\section{Introduction}

Graphics Processing Units (GPUs) have been effectively used to accelerate large data parallel applications. GPUs consisting of Streaming Multiprocessors (SMs) achieve high throughput by concurrently executing a large number of threads to hide long latencies. The throughput achieved by a GPU depends on the amount of thread level parallelism (TLP) utilized by it. Recent studies~\cite{NMNL}~\cite{ondemand}~\cite{WarpLevelDivergence}~\cite{SharedMemMultiplexing} focus on improving the throughput of GPUs by exploiting the TLP. 

The amount of TLP utilized by a GPU depends on the number of threads resident on it. When an application is launched on a GPU, an execution configuration consisting of the number of thread blocks and number of threads in a thread block is specified. The number of thread blocks that can actually be launched on an SM depends on the resource requirement, such as the number of registers and the amount of shared memory needed by each thread block. If an SM contains $R$ units of a resource and a thread block requires $R_{tb}$ units to complete its execution, then the SM can launch at the most $\left\lfloor{R}/{R_{tb}}\right\rfloor$ thread blocks, utilizing $R_{tb}\times\left\lfloor{R}/{R_{tb}}\right\rfloor$ units. The remaining $R \bmod R_{tb}$ units are wasted. 

In this paper we propose a mechanism for sharing of resources of SM in order to launch more thread blocks that
reduces the wastage of resources. In particular, we show how sharing of registers and sharing of scratchpad improves the throughput of SMs.
%
%
It is observed~\cite{NMNL} that increasing the number of threads benefits compute bound applications, but may result in increased L1/L2 cache misses for memory bound applications, thereby decreasing their performance. 
To overcome this, we propose an optimization, called \textit{Owner Warp First (OWF)} that schedules the extra thread blocks and their constituent warps effectively. For the register sharing approach, we further propose two optimizations, viz., \emph{Unrolling and Reordering of Register Declaration} and \emph{Dynamic Warp Execution} that improves register utilization and minimizes the number of stall cycles observed by the additional thread blocks respectively.

\begin{figure*}[t!]
  \renewcommand{\arraystretch}{.5}
  \begin{tabular}{@{}c@{}c@{}c@{}c}
    {\small \ \ \ \ \ (a)} \hspace*{-10mm}\includegraphics[scale=0.5]{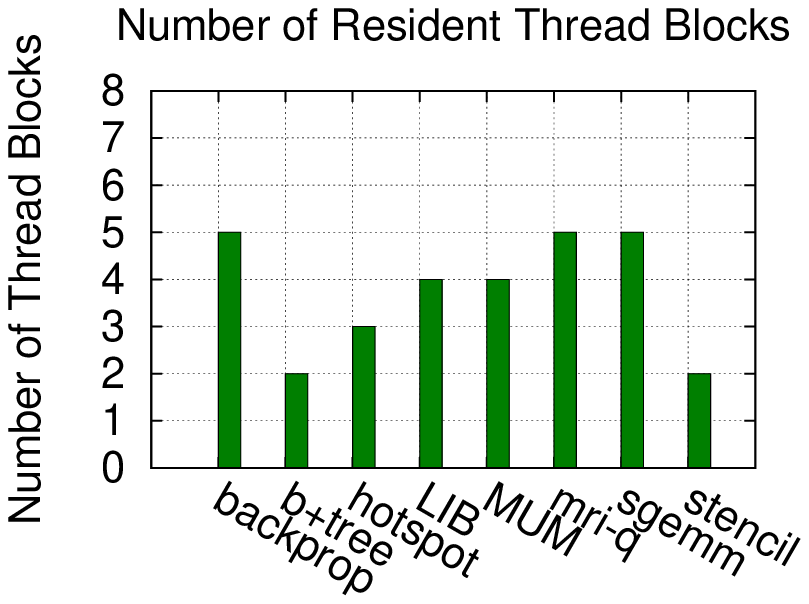} &
    {\small \ \ \ \ \ (b)} \hspace*{-10mm}\includegraphics[scale=0.5]{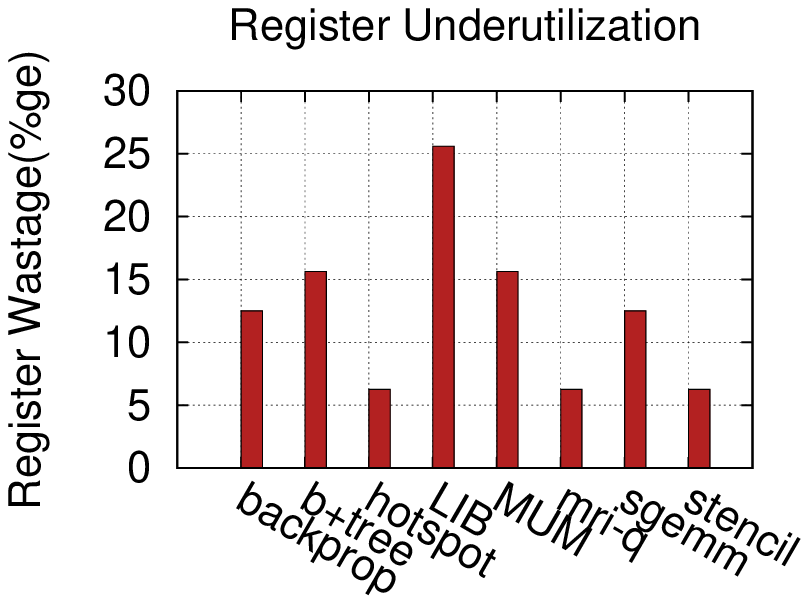} &
    {\small \ \ \ \ \ (c)} \hspace*{-10mm}\includegraphics[scale=0.5]{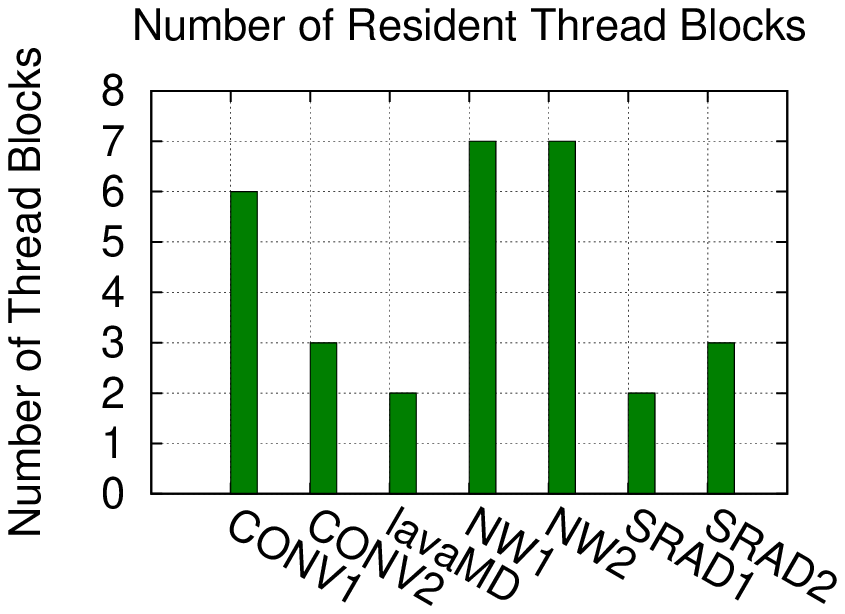} &
    {\small \ \ \ \ \ (d)} \hspace*{-10mm}\includegraphics[scale=0.5]{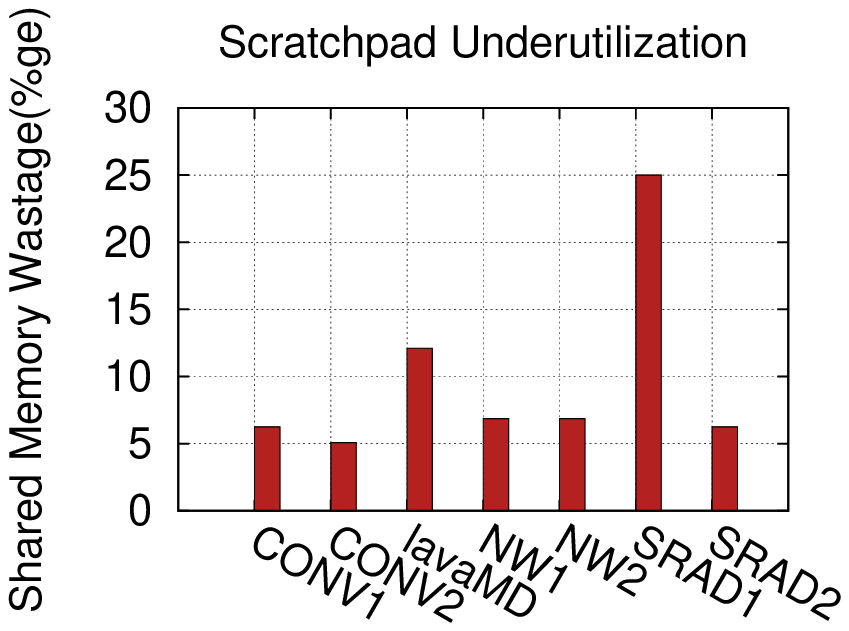}
  \end{tabular}        
\vspace*{-1mm}
  \caption{
    (a) Number of resident thread blocks with limited registers (b) Register underutilization \\
    (c) Number of resident thread blocks with limited scratchpad memory  (d) Scratchpad underutilization 
    \label{fig:motivation}}
\vspace*{-5mm}
\end{figure*}

\subsection{Motivation} \label{sec:motivation}


The problem of resource underutilization occurs in GPU, because resources are allocated at thread block granularity. We analyzed several benchmark applications using the GPGPU-Sim~\cite{GPGPUSIM} simulator\footnote{The GPU configuration is described in Table~\ref{table:GPGPUArch}, Section~\ref{sec:evaluation}. The benchmark details are given in  Table~\ref{table:set1}, Table~\ref{table:set2} Section~\ref{sec:evaluation}.}.  For applications that are limited by register resource, we show the number of resident thread blocks per SM in Figure~\ref{fig:motivation}(a), and we show the percentage of registers that are unutilized per SM in Figure~\ref{fig:motivation}(b). Consider the application \emph{hotspot}. Each thread for this benchmark needs 36 registers, and there are 256 threads in each block, so the number of registers required per thread block is 9216 (36 * 256).  According to the configuration (Table~\ref{table:GPGPUArch}), the number of registers available on an SM is 32768, so an SM can fit only 3 threads blocks ($\left\lfloor\frac{32768}{9216}\right\rfloor$). This results in 5120 registers per SM are wasted. 

Similarly, in Figure~\ref{fig:motivation}(c) we show the number of resident thread blocks per SM for the applications that are limited by scratchpad resource, and in Figure~\ref{fig:motivation}(d) we show the percentage of scratchpad memory that remains unutilized per SM.  Consider the application \emph{lavaMD}. Each thread block for this benchmark needs 7200 bytes of scratchpad memory. According to the configuration in Table~\ref{table:GPGPUArch}, the amount of scratchpad memory available per SM is 16384 bytes, hence an SM can fit only 2 threads blocks. This results in 1984 bytes of memory per SM remaining unutilized. Similar behavior is observed for other applications as well.

Applications that are constrained by their resource requirements may not only 
have low residency, but also waste resources of GPU. Our approach uses sharing to reduce the number of unutilized resources in order to increase the number of resident thread blocks. Our experiments show that these extra thread blocks help to hide long execution latencies and increase throughput.

\subsection{Contributions}\label{sec:contrib}
We make the following contributions in this work: 
\begin{myenumerate}
\item To utilize the resources of GPUs effectively, we propose a novel resource sharing mechanism that enables launching of  more thread blocks per SM.
\item We implemented our approach for two resources i.e, registers and  scratchpad. We propose optimizations to further improve the throughput of applications.
\item We implemented our approach using GPGPU-Sim simulator and evaluated on several applications from GPGPU-Sim, Rodinia, CUDA-SDK, and Parboil benchmark suites. We observed an average improvement of 11\% with register sharing and an average improvement of 12.5\% with scratchpad sharing.
\end{myenumerate}

The paper is organized as follows: 
Section~\ref{sec:background} describes the background required for our approach.  
Our approach is presented in Sections~\ref{sec:implementation} and~\ref{sec:optimizations}.
Section~\ref{sec:storage} discusses hardware overhead for implementing our approach.
Section~\ref{sec:experimentanalysis} describes the experimental evaluation.
Section~\ref{sec:relatedwork}  discusses related work and Section~\ref{sec:conclusion} concludes the paper.

\section{Background} \label{sec:background}
A typical NVIDIA GPU~\cite{CUDA} consists of a set of Streaming Multiprocessors~(SMs), and each multiprocessor has execution units called Stream Processors~(SPs). CUDA~\cite{CUDA} supports extensions to languages, such as C, to allow programmers to define and invoke parallel functions, called kernels, on a GPU. A kernel is invoked along with an execution configuration of threads that specifies the number of threads per thread block and the number of thread blocks.

The number of thread blocks that can reside on an SM depends on: (a) the number of registers used by a thread block and the number of registers available in the SM, (b) the amount of scratchpad memory used by a thread block and the amount of scratchpad memory available in the SM, (c) the maximum number of threads allowed 
per SM, and (d) maximum number of thread blocks allowed per SM. The threads in a thread block are
further divided into a set of consecutive 32 threads called Warp. Each SM 
contains one or more warp schedulers which schedule a ready warp every cycle
from a pool of ready warps. All threads in a warp execute the same instruction.

Warp schedulers schedule instructions in-order and so, when the current
instruction of a warp can not be issued, the warp is not considered to be ready.
If no warp can be scheduled in a cycle, then that is a stall cycle. As the
number of stall cycles increases, the run time goes up and the throughput 
decreases. Our approach increases the number of resident thread blocks by
utilizing the wasted registers as well as scratchpad memory on each SM and hence increases the number of resident warps and also improves the warp schedulers to hide long latencies.


\newcommand{\Warp}[2]{%
\psset{unit=.25mm}
\begin{pspicture}(0,0)(40,20)
\psframe(0,0)(40,20)
\putnode{zarb1342}{origin}{20}{10}{\rnode{#1}{#2}}
\end{pspicture}%
}

\begin{figure*}[t]
\renewcommand{\arraystretch}{0.5}
\begin{tabular}{@{}c@{\ \ \ }c@{\ \ \ }c@{}}
\scalebox{.75}{
\psset{unit=1mm}
\psset{linewidth=.3mm}
\begin{pspicture}(1,3)(63,48)
  \psframe(1,2)(63,48)
  \putnode{t0}{origin}{32}{45}{Number of Resource Units in SM = 35K}

  \psframe(5,9)(21,34)
  \putnode{o1}{origin}{13}{30}{\Warp{w0}{$w_0$}}
  \putnode{o2}{o1}{0}{-6}{\Warp{w1}{$w_1$}}
  \putnode{o3}{o2}{0}{-5}{\vdots}  
  \putnode{o9}{o3}{0}{-6}{\Warp{w9}{$w_9$}}
  \putnode{t1}{o1}{0}{8}{\begin{tabular}{c}TB$_0$\\(10K Units)\end{tabular}}
  \putnode{t2}{o9}{0}{-7}{\begin{tabular}{c}Unshared\end{tabular}}

  \psframe(24,9)(40,34)
  \putnode{o1}{origin}{32}{30}{\Warp{w0}{$w_{10}$}}
  \putnode{o2}{o1}{0}{-6}{\Warp{w1}{$w_{11}$}}
  \putnode{o3}{o2}{0}{-5}{\vdots}  
  \putnode{o9}{o3}{0}{-6}{\Warp{w9}{$w_{19}$}}
  \putnode{t1}{o1}{0}{8}{\begin{tabular}{c}TB$_1$\\(10K Units)\end{tabular}}
  \putnode{t2}{o9}{0}{-7}{\begin{tabular}{c}Unshared\end{tabular}}

  \psframe(43,9)(59,34)
  \putnode{o1}{origin}{51}{30}{\Warp{w0}{$w_{20}$}}
  \putnode{o2}{o1}{0}{-6}{\Warp{w1}{$w_{21}$}}
  \putnode{o3}{o2}{0}{-5}{\vdots}  
  \putnode{o9}{o3}{0}{-6}{\Warp{w9}{$w_{29}$}}
  \putnode{t1}{o1}{0}{8}{\begin{tabular}{c}TB$_2$\\(10K Units)\end{tabular}}
  \putnode{t2}{o9}{0}{-7}{\begin{tabular}{c}Unshared\end{tabular}}

\end{pspicture}}
&
\scalebox{.75}{
\psset{unit=1mm}
\psset{linewidth=.3mm}
\begin{pspicture}(1,3)(82,48)
  \psframe(1,2)(82,48)
  \putnode{t0}{origin}{41}{45}{Number of Registers (Reg) in SM = 35K}

  \psframe(5,9)(21,34)
  \putnode{o1}{origin}{13}{30}{\Warp{w0}{$w_0$}}
  \putnode{o2}{o1}{0}{-6}{\Warp{w1}{$w_1$}}
  \putnode{o3}{o2}{0}{-5}{\vdots}  
  \putnode{o9}{o3}{0}{-6}{\Warp{w9}{$w_9$}}
  \putnode{t1}{o1}{0}{8}{\begin{tabular}{c}TB$_0$\\(10K Reg)\end{tabular}}
  \putnode{t2}{o9}{0}{-7}{\begin{tabular}{c}Unshared\end{tabular}}

  \psframe(24,9)(40,34)
  \putnode{o1}{origin}{32}{30}{\Warp{w0}{$w_{10}$}}
  \putnode{o2}{o1}{0}{-6}{\Warp{w1}{$w_{11}$}}
  \putnode{o3}{o2}{0}{-5}{\vdots}  
  \putnode{o9}{o3}{0}{-6}{\Warp{w9}{$w_{19}$}}
  \putnode{t1}{o1}{0}{8}{\begin{tabular}{c}TB$_1$\\(10K Reg)\end{tabular}}
  \putnode{t2}{o9}{0}{-7}{\begin{tabular}{c}Unshared\end{tabular}}

  \putnode{o21}{origin}{51}{30}{\Warp{w0}{$w_{20}$}}
  \putnode{o22}{o21}{0}{-6}{\Warp{w1}{$w_{21}$}}
  \putnode{o23}{o22}{0}{-5}{\vdots}  
  \putnode{o29}{o23}{0}{-6}{\Warp{w9}{$w_{29}$}}
  \putnode{t1}{o21}{0}{10}{\begin{tabular}{c}TB$_2$\\\ \end{tabular}}
  \putnode{t1a}{t1}{-8}{-3}{}
  \putnode{t2a}{o29}{-8}{-7}{}

  \putnode{o31}{origin}{70}{30}{\Warp{w0}{$w_{30}$}}
  \putnode{o32}{o31}{0}{-6}{\Warp{w1}{$w_{31}$}}
  \putnode{o33}{o32}{0}{-5}{\vdots}  
  \putnode{o39}{o33}{0}{-6}{\Warp{w9}{$w_{39}$}}
  \putnode{t1}{o31}{0}{10}{\begin{tabular}{c}TB$_3$\\\ \end{tabular}}
  \putnode{t1b}{t1}{8}{-3}{}
  \putnode{t2b}{o39}{8}{-7}{}

  \ncline{|-|}{t2a}{t2b} \lput*{:U}{Shared}
  \ncline{|-|}{t1a}{t1b} \lput*{:U}{(15K Reg)}
  \psframe(62,9)(78,34)
  \psframe(43,9)(59,34)

  \ncline{<->}{o21}{o31}
  \ncline{<->}{o22}{o32}
  \ncline{<->}{o29}{o39}
\end{pspicture}}
&
\scalebox{.75}{
\psset{unit=1mm}
\psset{linewidth=.3mm}
\begin{pspicture}(1,3)(82,48)
  \psframe(1,2)(82,48)
  \putnode{t0}{origin}{41}{45}{Scratch Pad Memory (SPM) in SM = 35K}

  \psframe(5,9)(21,34)
  \putnode{o1}{origin}{13}{30}{\Warp{w0}{$w_0$}}
  \putnode{o2}{o1}{0}{-6}{\Warp{w1}{$w_1$}}
  \putnode{o3}{o2}{0}{-5}{\vdots}  
  \putnode{o9}{o3}{0}{-6}{\Warp{w9}{$w_9$}}
  \putnode{t1}{o1}{0}{8}{\begin{tabular}{c}TB$_0$\\(10K SPM)\end{tabular}}
  \putnode{t2}{o9}{0}{-7}{\begin{tabular}{c}Unshared\end{tabular}}

  \psframe(24,9)(40,34)
  \putnode{o1}{origin}{32}{30}{\Warp{w0}{$w_{10}$}}
  \putnode{o2}{o1}{0}{-6}{\Warp{w1}{$w_{11}$}}
  \putnode{o3}{o2}{0}{-5}{\vdots}  
  \putnode{o9}{o3}{0}{-6}{\Warp{w9}{$w_{19}$}}
  \putnode{t1}{o1}{0}{8}{\begin{tabular}{c}TB$_1$\\(10K SPM)\end{tabular}}
  \putnode{t2}{o9}{0}{-7}{\begin{tabular}{c}Unshared\end{tabular}}

  \putnode{o21}{origin}{50}{30}{\Warp{w0}{$w_{20}$}}
  \putnode{o22}{o21}{0}{-6}{\Warp{w1}{$w_{21}$}}
  \putnode{o23}{o22}{0}{-5}{\vdots}  
  \putnode{o29}{o23}{0}{-6}{\Warp{w9}{$w_{29}$}}
  \putnode{t1}{o21}{0}{10}{\begin{tabular}{c}TB$_2$\\\ \end{tabular}}
  \putnode{t1a}{t1}{-8}{-3}{}
  \putnode{t2a}{o29}{-8}{-7}{}

  \putnode{o31}{origin}{72}{30}{\Warp{w0}{$w_{30}$}}
  \putnode{o32}{o31}{0}{-6}{\Warp{w1}{$w_{31}$}}
  \putnode{o33}{o32}{0}{-5}{\vdots}  
  \putnode{o39}{o33}{0}{-6}{\Warp{w9}{$w_{39}$}}
  \putnode{t1}{o31}{0}{10}{\begin{tabular}{c}TB$_3$\\\ \end{tabular}}
  \putnode{t1b}{t1}{8}{-3}{}
  \putnode{t2b}{o39}{8}{-7}{}

  \ncline{|-|}{t2a}{t2b} \lput*{:U}{Shared}
  \ncline{|-|}{t1a}{t1b} \lput*{:U}{(15K SPM)}
  \psframe(64,9)(79,34)
  \psframe(42,9)(58,34)

  \putnode{s29}{o29}{1}{8}{}
  \putnode{s39}{s29}{19}{0}{}
  \ncline[nodesepA=17pt,nodesepB=15pt,doubleline=true]{<->}{s29}{s39}
\end{pspicture}}
\\ 
(a) & (b) & (c) 
\end{tabular}

\caption{ Approaches to Resource Allocation  (a) Default Approach (b)  Register Sharing (c) Scratch Pad Memory Sharing\label{fig:resourcealloc}}
\vskip -6mm
\end{figure*}
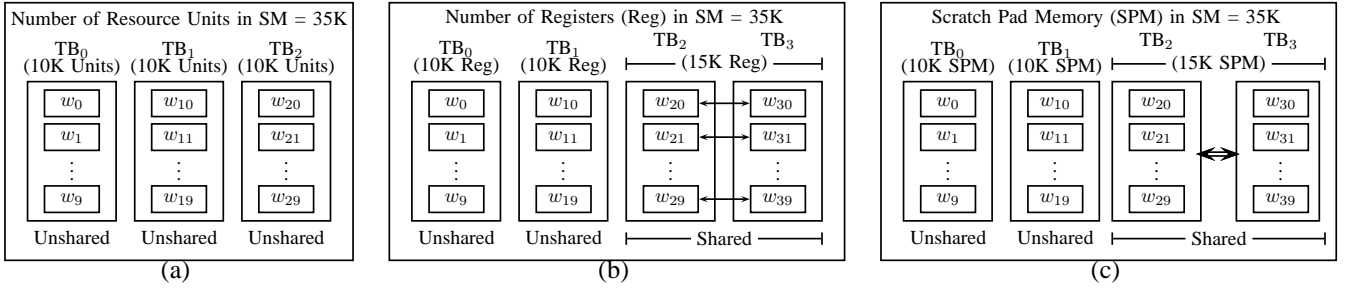

\section{Resource Sharing} \label{sec:implementation}

\newcommand{\TB}{\mbox{TB}}
We can increase the number of thread blocks in an SM by allowing two thread blocks to share resources. For example, consider an application that has thread blocks of size 10 warps (320 threads), and a thread block requires 10K resource units to complete its execution. If an SM has 35K resource units, at most 3 thread blocks can be resident on each SM by utilizing 30K resource units; the remaining 5K units are wasted. The schematic of this approach (baseline) is shown in Figure~\ref{fig:resourcealloc}(a), where thread blocks $\TB_0, \TB_1$ and $\TB_2$ are scheduled on an SM. 

In order to reduce the wastage of resources, our approach allocates one more thread block ($\TB_3$) in {\em sharing} mode with $\TB_2$. Instead of allocating 10K resource units separately to each of the thread blocks $\TB_2$ and $\TB_3$, a total of 15K units for the two blocks are allocated as follows: each of $\TB_2$ and $\TB_3$ is allocated 5K units exclusively ({\em Private or Unshared Resource}), while the remaining 5K units ({\em Shared Resource}) are all allocated to $\TB_2$ or $\TB_3$ whoever needs {\em any one of these resource first}. The thread block which did not get the ownership of shared resources waits when it needs {\em any} of the shared resource till the other block finishes.


We now describe in detail our approach for two types of resources (a) Register sharing (b) Scratchpad sharing.

\subsection{Register Sharing}
The scenario in Figure~\ref{fig:resourcealloc}(a) can be improved using our register allocation scheme shown in Figure~\ref{fig:resourcealloc}(b), in which we allocate 10K registers to each thread blocks $\TB_0$ and $\TB_1$. The remaining 15K registers are shared between thread blocks $\TB_2$ and $\TB_3$ such that each pair of warps in these thread blocks are allocated 1.5K registers as described next. We refer to TB0 and TB1 as unshared thread blocks, whereas, TB2 and TB3 as shared thread blocks.

Consider the pair of warps $W_{20}$ and $W_{30}$ that participate in sharing. Our approach allocates 0.5K registers each to $W_{20}$ and $W_{30}$ exclusively (private or unshared registers). The remaining 0.5K registers are shared registers, that are allocated to these warps {\em together} in a shared but exclusive manner, i.e. only one of them can access the pool of shared registers at a time. For example, if warp $W_{20}$ accesses any of the shared registers first, exclusive access to all the 0.5K shared registers is given to $W_{20}$, while $W_{30}$ is prevented from accessing any of those 0.5K shared registers till $W_{20}$ finishes. This implies, $W_{30}$ can continue its execution until its first access to any of the 0.5K shared registers, at which point it busy-waits. Only after $W_{20}$ finishes execution, $W_{30}$ can access the shared registers and continue. This way, additional warps make {\em some} progress, which helps in hiding execution latencies. 

To generalize this idea and to compute the increase in number of thread blocks, we will consider a GPU that provides $R$ registers per SM. Also, consider a thread block that requires $R_{tb}$ registers, and each warp in the thread block requires $R_w$ registers to complete its execution. 
To increase the number of thread blocks that share registers with other existing thread blocks in the SM, we allocate $R_{tb}(1+t)$ (for any threshold 0 $<$ t $\leq$ 1) registers to each pair of shared thread blocks, instead of allocating $2R_{tb}$ registers to them (in Figure~\ref{fig:resourcealloc}(b), $t$ is 0.5).  
Equivalently we allocate $R_w(1 + t)$ registers per two warps from these thread blocks (i.e, one warp from each shared thread block in the pair), such that each of these warps can access $R_wt$ unshared registers independently, and they can access the remaining $R_w(1-t)$ shared registers only when granted access.

\newcommand{\Alu}[1]{%
\scalebox{.85}{
  \psset{unit=1mm}
  \begin{pspicture}(0,0)(12,6)
  \pspolygon(2,0)(10,0)(12,6)(0,6)
  \rput(6,3){#1}
  \end{pspicture}}%
}

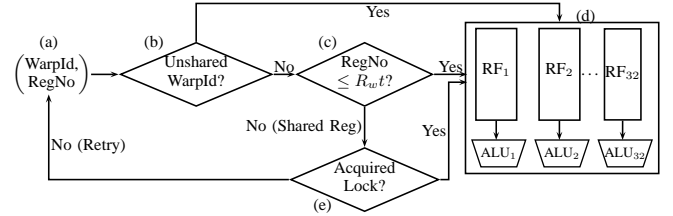
\begin{figure}[t]
\scalebox{.70}{\small
  \psset{unit=1mm}
  \psset{linewidth=.3mm}
  \begin{pspicture}(0,0)(120,40)
    \psframe(86,6)(123,35)
    \putnode{t0}{origin}{7}{25}{
      $\left(\begin{tabular}{@{}c@{}}
        WarpId,\\ RegNo
    \end{tabular}\right)$}
    
    \putnode{t1}{t0}{28}{0}{\psdiabox[framesep=-3]{
        \rule[-4.5mm]{0cm}{10mm}\begin{tabular}{c}
          Unshared\\WarpId?
        \end{tabular}
    }}
    
    \putnode{t2}{t1}{32}{0}{\psdiabox[framesep=-3]{
        \rule[-4.5mm]{0cm}{10mm}\begin{tabular}{c}
          RegNo \\ $\leq R_wt$?
        \end{tabular}
    }}

    \putnode{t3}{t2}{0}{-20}{\psdiabox[framesep=-3]{
        \rule[-4.5mm]{0cm}{10mm}\begin{tabular}{c}
          Acquired \\ Lock?
        \end{tabular}
    }}
    
    \putnode{t4}{t2}{25}{0}{\psframebox{\rule[-7.5mm]{0mm}{15mm}RF$_1$}}
    \putnode{t5}{t4}{12}{0}{\psframebox{\rule[-7.5mm]{0mm}{15mm}RF$_2$}}
    \putnode{t6}{t5}{6}{0}{\ldots}
    \putnode{t7}{t6}{6}{0}{\psframebox[framesep=0]{\rule[-8mm]{0mm}{17mm}RF$_{32}$}}
    
    \putnode{a4}{t4}{0}{-15}{\Alu{ALU$_1$}}\ncline{->}{t4}{a4}
    \putnode{a5}{t5}{0}{-15}{\Alu{ALU$_2$}}\ncline{->}{t5}{a5}
    \putnode{a7}{t7}{0}{-15}{\Alu{ALU$_{32}$}}\ncline{->}{t7}{a7}

    \ncline{->}{t0}{t1}
    \ncline{->}{t1}{t2}\Aput[.2]{No}
    \ncline[nodesepB=1.5]{->}{t2}{t4}\Aput[.2]{Yes}
    \ncline{->}{t2}{t3}\Bput[.2]{No (Shared Reg)}
    \ncangle[angleA=180,angleB=-90]{->}{t3}{t0}\Bput[.2]{No (Retry)}
    \ncangle[angleA=90,angleB=90,nodesepB=1.1]{->}{t1}{t5}\Bput[.2]{Yes}
    \ncangle[angleA=0,angleB=180,nodesepB=1.5,offsetB=-1.5]{->}{t3}{t4}\Aput[.2]{Yes}

    \putnode{ca}{t0}{0}{6}{(a)}
    \putnode{cb}{t1}{-8}{6}{(b)}
    \putnode{cc}{t2}{-7}{6}{(c)}
    \putnode{cd}{t5}{5}{11}{(d)}
    \putnode{ce}{t3}{-8}{-5}{(e)}
\end{pspicture}}
\caption{Register Access Mechanism\label{fig:regaccess}}
\vskip -2mm
\end{figure}

\begin{figure}[t]
\scalebox{.70}{\small
  \psset{unit=1mm}
  \psset{linewidth=.3mm}
  \begin{pspicture}(0,0)(120,40)
    \putnode{t0}{origin}{8}{25}{
      $\left(\begin{tabular}{@{}c@{}}
        ThId,\\ SMemLoc
    \end{tabular}\right)$}
    
    \putnode{t1}{t0}{35}{0}{\psdiabox[framesep=-3]{
        \rule[-6mm]{0cm}{12mm}\begin{tabular}{c}
          ThId $\in$ \\ Unshared \\ ThreadBlock?
        \end{tabular}
    }}
    
    \putnode{t2}{t1}{45}{0}{\psdiabox[framesep=-3]{
        \rule[-4.5mm]{0cm}{10mm}\begin{tabular}{c}
          SMemLoc \\ $\leq R_{tb}t$?
        \end{tabular}
    }}

    \putnode{t3}{t2}{0}{-20}{\psdiabox[framesep=-3]{
        \rule[-4.5mm]{0cm}{10mm}\begin{tabular}{c}
          Acquired \\ Lock?
        \end{tabular}
    }}
    
    \putnode{t4}{t2}{27}{0}{
      \begin{tabular}[b]{|c|}\hline
        \rule{3mm}{0mm}\\\hline
        \\\hline
        \\\hline
        \\\hline
        \\\hline
      \end{tabular}\rotateleft{Scratchpad}
    }
    
    \putnode{a4}{t4}{0}{-15}{\Alu{ALU}}\ncline{->}{t4}{a4}

    \ncline{->}{t0}{t1}
    \ncline{->}{t1}{t2}\Aput[.2]{No}
    \ncline{->}{t2}{t4}\Aput[.2]{Yes}
    \ncline{->}{t2}{t3}\Bput[.2]{No (Shared Loc)}
    \ncangle[angleA=180,angleB=-90]{->}{t3}{t0}\Bput[.2]{No (Retry)}
    \ncangle[angleA=90,angleB=90]{->}{t1}{t4}\Bput[.2]{Yes}
    \ncangle[angleA=0,angleB=180,offsetB=-1.5]{->}{t3}{t4}\Aput[.2]{Yes}

    \putnode{ca}{t0}{0}{6}{(a)}
    \putnode{cb}{t1}{-9}{6}{(b)}
    \putnode{cc}{t2}{-7}{6}{(c)}
    \putnode{cd}{t4}{-8}{9}{(d)}
    \putnode{ce}{t3}{-8}{-5}{(e)}
\end{pspicture}}
\caption{Scratchpad Access Mechanism\label{fig:spaccess}}
\vskip -4mm
\end{figure}
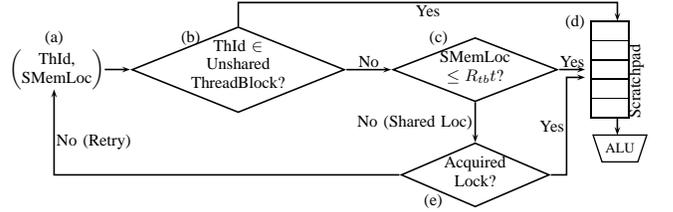

To detect a register accessed by a warp as shared or unshared, and to efficiently access it from the register file unit, we modify the existing register file access mechanism as shown in Figure~\ref{fig:regaccess}. When a warp (WarpId) needs to access a register (RegNo), we first check if the warp is an unshared warp i.e. if it belongs to an unshared thread block (Figure~\ref{fig:regaccess}, Step~(b)). If it is an unshared warp, it can directly access the register from register file using a combination of (WarpId, RegNo). If WarpId is a shared warp, the accessed register is an unshared register if $\mbox{RegNo} \leq R_wt$ (Step~(c)). This is because $R_wt$ number of unshared registers are allocated to each warp. If  $\mbox{RegNo} > R_wt$, we treat the register as a shared register. A warp can access an unshared register directly from the register file, but it can access a shared register only when it gets exclusive access by acquiring a lock (Step~(e)), otherwise it retries the access in another cycle. 

\newcommand{\Temp}[2]{%
\psset{unit=.25mm}
\begin{pspicture}(0,0)(65,30)
\psframe(0,0)(65,30)
\putnode{zarb1342}{origin}{31}{15}{\rnode{#1}{#2}}
\end{pspicture}%
}

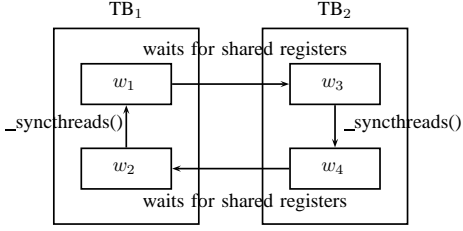
\begin{figure}[t]
\begin{tabular}{@{}c}
\scalebox{0.75}{
\psset{unit=1mm}
\psset{linewidth=.3mm}
\begin{pspicture}(45,20)(138,70)
	\psframe(68,26)(94,61) 
	\psframe(105,26)(131,61	) 
	\putnode{o1}{origin}{81}{51}{\Temp{w1}{$w_1$}}
	\putnode{o2}{o1}{0}{-15}{\Temp{w2}{$w_2$}}
	\putnode{o3}{o1}{37}{0}{\Temp{w3}{$w_3$}}
	\putnode{o4}{o3}{0}{-15}{\Temp{w4}{$w_4$}}    
	\ncline{->}{o1}{o3}
	\ncline{->}{o2}{o1}
	\ncline{->}{o4}{o2}
	\ncline{->}{o3}{o4}
	\putnode{t1}{o9}{49}{51}{\begin{tabular}{c}TB$_1$\end{tabular}}	
	\putnode{t1}{o9}{86}{51}{\begin{tabular}{c}TB$_2$\end{tabular}}		
	\putnode{t1}{o9}{70}{44}{\begin{tabular}{c}waits for shared registers\end{tabular}}			
	\putnode{t1}{o9}{70}{17}{\begin{tabular}{c}waits for shared registers\end{tabular}}			
	\putnode{t1}{o9}{38}{31}{\begin{tabular}{c}\_syncthreads()\end{tabular}}			
	\putnode{t1}{o9}{98}{31}{\begin{tabular}{c}\_syncthreads()\end{tabular}}			
	
\end{pspicture}} 
\end{tabular}
\vskip -4mm
        \caption{ Deadlock in the presence of barrier instructions \label{fig:deadlock}}
\vskip -5mm
\end{figure}


Consider a scenario shown in Figure~\ref{fig:deadlock}, where two thread blocks $\TB_1$ and $\TB_2$ are in shared mode. Assume that warp $W_1$ of $\TB_1$ tries to acquire a lock to access shared registers, but $W_3$ has already acquired the lock. Also, assume that the warps $W_2$, $W_3$ are waiting for warps $W_1$ and $W_4$ respectively, to arrive at a barrier instruction (\_syncthreads()). 
Now, if warp $W_4$ tries to acquire a lock to access shared registers from $W_2$, and $W_2$ has already a lock, then a deadlock occurs. To avoid deadlock, we always ensure that if thread blocks $\TB_1$ and $\TB_2$ share registers, then a warp from $\TB_1$ can acquire a lock only when either (a) none of the warps from $\TB_2$ have acquired a lock for the shared registers, or (b) the warps from $\TB_2$ that have acquired exclusive access to the shared registers have finished their execution. For the example, since warp $W_2$ already has acquired a lock, $W_3$ can not acquire a lock, hence avoiding the deadlock.

\subsection{Scratchpad Sharing}

Figure~\ref{fig:resourcealloc}(c) shows an example of \textit{Scratchpad Sharing}, where we consider a GPU that has 35K units of scratchpad memory per SM, and each thread block requires 10K units. To increase number of thread blocks, we allocate 10K units to each $\TB_0$ and $\TB_1$. We allocate the remaining 15K scratchpad units together for thread blocks $\TB_2$ and $\TB_3$.  With scratchpad sharing, each of the thread block $\TB_2$ and $\TB_3$ can access 5K scratchpad memory each privately. Remaining 5K units of memory is allotted to the first block that accesses it\footnote{Unlike register sharing, we can not distribute 1.5K scratchpad memory to each pair of warps, because any thread within a thread block can access any scratchpad location allocated for that block.}. Similar to Register sharing approach, we refer to TB0 and TB1 as unshared thread blocks, whereas, TB2 and TB3 as shared thread blocks.
When a thread from the shared thread block (say $\TB_2$) needs to access a memory location from shared scratchpad, it gains an exclusive access by acquiring a lock. Any thread from the other shared thread block ($\TB_3$) can continue to execute instructions till the first access of shared scratchpad location, at which point all threads of $\TB_3$ busy-wait till $\TB_2$ has finished its execution.

The implementation to support scratchpad sharing in GPGPU-Sim is shown in Figure~\ref{fig:spaccess}. 

 The steps for the shared scratchpad access follow the rules similar to the shared register access, and are omitted for brevity. Note that a deadlock can never occur with scratchpad sharing. 

\subsection{Computing No of Thread Blocks to be Launched per SM} \label{Max_thread block}

A naive method of sharing, where each thread block is sharing resources with some other thread block, may launch more thread blocks as compared to default (non-sharing) approach. However, the number of thread blocks  that make progress ({\em effective thread blocks}) per SM can be less than that for non-sharing. For example, consider a scenario where 3 thread blocks are resident per SM without sharing. With naive sharing approach, it may be possible to have 4 thread blocks resident, such that block 1 shares resources with block 2; and block 3 shares resources with block 4. It can happen that block 2 and 4 start accessing shared resources causing blocks 1 and 3 to wait. Effectively only two thread blocks (blocks 2 and 4) will make progress in the naive sharing approach, whereas all 3 blocks can make progress in the non-sharing approach reducing the throughput. To avoid this,  we describe a method to compute the total number of thread blocks (Shared + Unshared) to be launched per SM such that the number of effective thread blocks using sharing is no less than that of non-sharing approach. We use the following notations:
\begin{myenumerate}
\item $R$: Number of units of resource available per SM,
\item $R_{tb}$: Number of resource units required by a thread block,
\item $S$: Number of pairs of thread blocks that are to be launched per SM in shared mode,
\item $U$: Number of thread blocks to be launched in an SM that do not share resources with any other thread block,
\item $M$: Maximum number of thread blocks to be launched in an SM,
\item $t$: Threshold for computing the number of resources that a thread block shares with another thread block. For a given threshold value $t$ ($0 < t \leq 1$) we allocate $(1+t)R_{tb}$ resource units per two shared thread blocks, in which we use $(1-t)R_{tb}$ resource units as shared units.
\end{myenumerate}

Without sharing, we can launch up to
$\left\lfloor{R}/{R_{tb}}\right\rfloor$ thread blocks in an SM, and all of them make progress. Whereas in our approach, if two thread blocks are launched in sharing mode, at least one thread block always makes progress. So, when $S$ shared pairs are launched in an SM, at least $S$ thread blocks always make progress. Also, if $U$ unshared thread blocks are launched in the SM, they always make progress. Therefore, at least $S+U$ thread blocks always make progress with our approach. 
In order to keep the number of effective thread blocks in our approach to be same as that of no-sharing approach, we need the following relation to hold:
\begin{equation}\label{eq:max_unblock}
S+U =  \left\lfloor{R}/{R_{tb}}\right\rfloor
\end{equation}
For each shared pair of thread blocks, we allocate $R_{tb}(1+t)$ resource units. Similarly for each unshared thread block, we allocate $R_{tb}$ resource units. Since the total number of resource units available in the SM is $R$, we have:
\begin{equation} \label{eq:tot_reg}
UR_{tb}+ SR_{tb}(1+t) \leq R 
\end{equation}
 The total number of thread blocks that can be launched in sharing approach is equal to the number of unshared thread blocks plus twice the number of shared pairs, i.e,
\begin{equation}\label{eq:max_cta}
M = U + 2S 
\end{equation}
Using Equations~\ref{eq:max_unblock},~\ref{eq:tot_reg}, and~\ref{eq:max_cta},
\begin{equation}
M  =  \left\lfloor\dfrac{R}{R_{tb}}\right\rfloor\ +\ \frac{1}{t}\left(\frac{R}{R_{tb}}\ -\ \left\lfloor\dfrac{R}{R_{tb}}\right\rfloor\right)
\end{equation}

Since the actual number of thread blocks that can reside in an SM also depends on (a) maximum number of resident threads per SM, and (b) maximum number of resident thread blocks in the SM; the number of thread blocks resident in an SM in our approach is the minimum of numbers obtained using these factors and the value of $M$.

\newcommand{\SM}{\mbox{SM}}
\section{Optimizations} \label{sec:optimizations}

\begin{figure}
\includegraphics[scale=0.45]{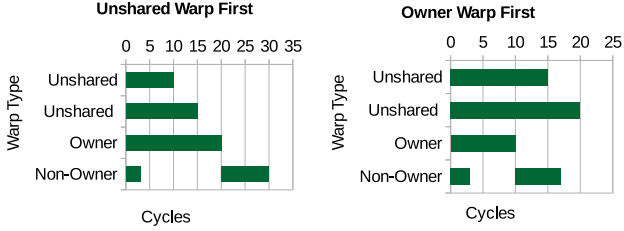}
\vskip -3mm
\caption{Warp Scheduling}
\label{fig:OWF}
\vskip -5mm
\end{figure}

With the proposed resource sharing approach, each SM has unshared and shared warps, and scheduling these warps plays a very important role in determining the performance of applications. We propose an optimization called ``Owner Warp First (OWF)" to schedule these warps effectively. If two thread blocks $\TB_i$ and $\TB_j$ are a shared pair, and at least one of the warps of $\TB_i$ waits for shared resources from $\TB_j$, we call $\TB_j$ as {\em Owner Block}, and the warps that belong to $\TB_j$ are called {\em Owner Warps}. $\TB_i$ is called {\em Non-Owner Block} and warps of $\TB_i$ are called {\em Non-Owner Warps}. As soon as the owner thread block finishes its execution, it transfers its ownership to the non-owner thread block (i.e, the non-owner thread block becomes the owner), and a new non-owner thread block gets launched.

\subsection{Scheduling Owner Warp First (OWF)} 

A warp scheduler in the SM issues a warp every cycle from a pool of ready warps.
With our solution, the warps can be categorized into three types viz., unshared, shared owner and shared non-owner. In register sharing, shared non-owner warps depend on the corresponding shared-owner warps to release registers, before they can make progress. Similarly with scratchpad sharing, warps from non-owner thread blocks wait for owner thread blocks to complete their execution. In Owner Warp First algorithm we prioritize warps in the order: shared owner, unshared, and shared non-owner. Giving the highest priority to shared owner warps helps finish them sooner, and hence the dependent shared non-owner warps can make progress. Since shared non-owner warps can not
make much progress before stalling, giving them lower priority than unshared warps helps use them only to hide stalls when no other types of warps are ready to run.
Figure~\ref{fig:OWF} illustrates that scheduling unshared warps with high priority compared to shared warps can degrade the performance of an application. 

\begin{figure}[t]
 \scalebox{.9}{\begin{tabular}{@{}c@{}|@{}c@{}}
  \texttt{\renewcommand{\arraystretch}{0.5}
  \begin{tabular}[t]{@{}l@{}}
    .reg .u32 \$r\textless 27\textgreater ; \\ 
    .reg .u32 \$ofs\textless 3\textgreater; \\ 
    .reg .pred \$p\textless 4\textgreater; \\
    .reg .u32 \$r124; \\  
    .reg .u32 \$o127; \\ 
    \\    \\    \\    \\    \\ \\ \\
    set.le.s32.s32 \$p0/\$o127, \\
    $\qquad\qquad$s[0x003c], \$r124; \\ 
    mov.u32 \$r16, \$r124; \\
    mov.u32 \$r17, \$r124; \\
    mov.u32 \$r9, \$r124; \\
    mov.u32 \$r18, \$r124; \\
    mov.u32 \$r10, \$r124; \\ 
    /* Code here */  \\
  \end{tabular}}
&
  \texttt{\renewcommand{\arraystretch}{0.5}
  \begin{tabular}[t]{l}
    .reg .pred \$p0; \\
    .reg .u32 \$o127;	\\
    .reg .u32 \$r124;\\
    .reg .u32 \$r16;\\
    .reg .u32 \$r17;\\
    .reg .u32 \$r9;\\
    .reg .u32 \$r18; \\	
    .reg .u32 \$r10;\\
    .reg .u32 \$r11;\\
    \ldots \\
    set.le.s32.s32 \$p0/\$o127,\\
    $\qquad\qquad$s[0x003c], \$r124; \\ 
    mov.u32 \$r16, \$r124; \\
    mov.u32 \$r17, \$r124; \\
    mov.u32 \$r9, \$r124; \\
    mov.u32 \$r18, \$r124; \\
    mov.u32 \$r10, \$r124; \\ 
    /* Code here */  \\
  \end{tabular}} \\ 
  (a) Normal Declarations & (b) Unrolled Declarations
\end{tabular}}
  \caption{ Unrolling and Reordering of Register Declarations \label{fig:unroll}}
\vskip -5mm
\end{figure}


\subsection{Unrolling and Reordering of Register Declarations} 
In register sharing, non-owner warps need to wait for owner warps when they try to access shared registers. If the very first instruction issued by a non-owner warp uses a shared register, then the warp has to wait and can not start its execution until corresponding owner warp has released the shared register. In order to allow the non-owner warps to execute as many instructions as possible before stalling due to
unavailability of shared registers, we unroll and reorder the register declarations.
To illustrate this, consider the PTXPlus~\cite{GPGPUSIM} code shown in Figure~\ref{fig:unroll}(a), which is generated by GPGPU-Sim~\cite{GPGPUSIM} for the \textit{sgemm} application from Parboil Suite~\cite{parboil}. The first instruction of the code accesses registers p0 and r124, which get the register sequence numbers as 31 and 35 according to the declaration. These registers are part of the shared registers for a certain threshold value t. Hence, a non-owner warp has to wait until the registers are released. To delay accessing the shared registers, we unroll and rearrange the order of the register declarations so that p0, r124 become unshared registers (i.e, they get the register sequence numbers as 1 and 3). This is shown in Figure~\ref{fig:unroll}(b). Hence the non-owner warps get to execute more number of instructions before they start accessing shared registers.

To implement this optimization, we converted the assembly code (PTXPlus) produced by GPGPU-Sim into an optimized assembly code. To achieve this, we first find an order of registers according to their first usage. Further, to ensure that the unshared registers are used before the shared registers, we modify the register declarations so that a register that has been used first in assembly code is declared first. Finally, we modified the GPGPU simulator to use the optimized PTXPlus code for simulating instructions. This optimization can be easily integrated at the assembly level using the existing CUDA compiler.

\subsection{Dynamic Warp Execution} 
Recent study by Kayiran et. al.~\cite{NMNL} shows that the performance of memory bound applications can degrade with increase in the number of resident thread blocks. Executing additional thread blocks can  increase L1/L2 cache misses, which leads to increase in the number of stall cycles. Register sharing approach exploits more thread level parallelism from the additional warps (non-owner warps) compared to scratchpad sharing, because register sharing allows non-owner warps to start executing as soon as its corresponding owner warp finishes. So, in order to reduce the number of stalls raised due to execution of non-owner warps, we propose an optimization that can dynamically enable or disable the execution of long latency instructions (memory) issued by the additional blocks.

To control the execution of memory instructions from the non-owner warps, we monitor the number of stall cycles for each SM. When executing memory instructions from non-owner warp leads to increase in the number of stalls, we decrease the probability of executing further memory instructions from the non-owner warps. To illustrate this, consider a GPU that has N SMs, all in sharing mode. Our approach disables execution of memory instructions for the non-owner warps, only on a specific SM (e.g. $\SM_0$). Every other SM, $\SM_i$ for $i \in \{1\ldots N-1$\}, allows execution of memory instructions for the non-owner warps, and compares its stall cycles periodically with the stalls on $\SM_0$.  If stalls observed in the $\SM_i$ are more than the stalls appearing in $\SM_0$, then the probability of executing memory instructions on $\SM_i$ from the non-owner warps is decreased by a predetermined value $p$. If the stalls in $\SM_i$ are less than that in $\SM_0$, then the probability of executing memory instructions on $\SM_i$ from the non-owner warps is increased by the same value $p$. Thus, we reduce the number of stall cycles by controlling the execution of memory instructions.

After running several experiments, we selected the periodicity of monitoring to be 1000 cycles, which is to ensure that (a) the monitoring overhead is not high, and (b) sufficient number of stall cycles are observed.  In our experiments, initially all the $\SM$s are allowed to execute all memory instructions, i.e., the probability of executing memory instructions from non-owner warp is 1. Depending on the stall cycles observed for an $\SM_i$ ($i \in \{1\ldots N-1$\}), this probability for $\SM_i$ is decreased or increased by $p=0.1$, but is kept within interval $[0-1]$ as a saturating counter.

\section{Hardware Requirement} \label{sec:storage}
For both register and scratchpad sharing: (1) Each SM requires a
bit to specify whether sharing mode is enabled for it. This
bit will be set when the number of thread blocks assigned to
the SM using resource sharing is more than the default number
of thread blocks per SM. (2) Each thread block stores id of the
partner thread block (set to -1 if the thread block is in
unsharing mode). For T thread blocks, $T\lceil\log_2 (T+1)
\rceil$ (Assuming ids 0 to T-1 for T thread blocks, we can
use id T to represent -1) bits are required per SM. (3) Each warp requires a bit
for specifying the owner information. This bit is set only
when the warp is an owner warp.  Hence for W warps, W bits
are needed.

For register sharing: (1) Each warp requires a bit to specify
whether it is in sharing or unsharing mode.  Hence for $W$
warps in an SM, $W$ bits are required. For a warp in shared
mode, its corresponding shared warp can be identified using
the sharer thread block id of its thread block and its
relative position in the thread block.  (2) Each pair of shared
warps uses a lock variable to access the shared registers
exclusively. The lock variable is set to the id of the warp
which has gained access to the shared registers.  If an SM
has $W$ warps, there can be a maximum of
$\left\lfloor{W}/{2}\right\rfloor$ shared pairs of warps
in the SM. Hence we need a total of
$\left\lfloor{W}/{2}\right\rfloor \lceil\log_2 W\rceil$
bits per SM.

For scratchpad sharing: (1) Each pair of shared thread blocks
uses a lock variable to access the shared locations
exclusively. The lock variable is set to the id of the thread
block which has gained access to the shared scratchpad
region. If an SM has T thread blocks, there can be a maximum
of $\left\lfloor{T}/{2}\right\rfloor$ shared pairs of
thread blocks in the SM. Hence we need a total of
$\left\lfloor{T}/{2}\right\rfloor \lceil\log_2 T\rceil$
bits per SM.

Hence the total amount of storage required (in bits) for a GPU with $N$ SMs for
implementing register sharing is:
\[ (1+T\lceil\log_2 (T+1)
\rceil+2W+\left\lfloor{W}/{2}\right\rfloor \lceil\log_2 
W\rceil)*N\]
and for implementing  scratchpad sharing is: \[(1+T\lceil\log_2 (T+1)
\rceil+W+\left\lfloor{T}/{2}\right\rfloor \lceil\log_2 
T\rceil)*N\]
We also require two comparator circuits to implement
the resource accesses (Figures~\ref{fig:regaccess}
and~\ref{fig:spaccess}, steps~(b) and (c)).

\section{Experimental Analysis}\label{sec:experimentanalysis} 

\subsection{Evaluation Methodology} \label{sec:evaluation} 
We implemented our approach using GPGPU-Sim V3.X [8]. The baseline architecture used for comparing our
approach is shown in Table I. We experimentally evaluated our approach on several applications from GPGPU-Sim [8], Rodinina [9], CUDA-SDK [10], and Parboil [7] benchmarks. Depending on the resource requirement of applications, we divided the benchmarks in three sets. Set-1, shown in Table II, consists of applications whose number of thread blocks per SM is limited by the registers. Similarly Set-2, as show in Table III, has applications that are limited by scratch pad memory. Finally Set-3, shown in Table IV, consists of applications whose number of thread blocks per SM is limited by factors other than registers, such as (a) maximum number of resident threads (b) maximum number of resident thread blocks. For each application in the Tables II, III, and IV, we show names of the kernels used for evaluation, number of threads per thread block. In Table II, we report the the number of registers per thread for each kernel, which GPGPU-Sim uses to compute the number of resident thread blocks, and in Table III we show the amount scratch pad memory used by each thread block. We compiled all the applications using CUDA 4.2 and executed on the inputs provided in the benchmark suites.

We use the value of threshold ($t$) to configure the percentage of resource sharing. For example, if each thread block requires $R_{tb}$ units of resource, and we choose $t=0.1$, then we allocate $1.1*R_{tb}$ resource units per two shared thread blocks, which means 90\% of the resource units ($R_{tb}$) are used as shared resource units. So for a given threshold t, we can compute the percentage of register sharing as $(1-t)*100$. For all our experimental results, we use
the threshold value as 0.1 (i.e, 90\% resource sharing), unless otherwise specified.

\begin{table}[t]
\caption{GPGPU-Sim Architecture}
\vskip -1mm
\scalebox{0.90}{\renewcommand{\arraystretch}{.8}
\begin{tabular}{@{}l@{\ }l@{}}
\hline\hline
Resource & GPU Configuration \\
\hline
Number of Clusters & 14 \\
Number of Cores/Cluster & 1 \\
Max Number of Thread Blocks/Core & 8 \\
Max Number of Threads/Core & 1536  \\ 
Number of Registers/Core \space \space \space & 32768  \\ 
Scratchpad Memory/Core & 16KB \\
Warp Scheduling & LRR \\
Number of Schedulers & 2 \\
L1-Cache/Core & 16KB \\
L2-Cache & 768KB \\
DRAM Scheduler & FR-FCFS  \\
GDDR3 Timings & 
$\begin{array}[t]{@{}l@{}} 
  t_{RRD}=6, t_{WR}=12, 
  t_{RCD}=12,  t_{RAS}=28, \\
  t_{RP}=12, t_{RC}=40, 
  t_{CL}=12,t_{CDLR}=5  
\end{array}$ \\
\hline
\end{tabular}}
\label{table:GPGPUArch}
\vskip -1mm
\end{table}

\begin{table}[t]
\caption{Set-1: Benchmarks that are limited by registers}
\vskip -1mm
\renewcommand{\arraystretch}{.8}
\begin{tabular}{@{}l@{\ }l@{\ }l@{\ }l@{\ }l@{}}
\hline\hline
Benchmark &  Application & Kernel & Block & Registers \\
          &              &        & Size  & per thread \\
\hline
GPGPU-Sim  & backprop & bpnn\_adjust\_weights\_cuda & 256 & 24 \\
GPGPU-Sim  & b+tree & findRangeK & 508 & 24 \\
RODINIA & hotspot & calculate\_temp & 256 & 36 \\
RODINIA & LIB & Pathcalc\_Portfolio\_KernelGPU & 192& 36 \\
RODINIA & MUM & mummergpuKernel & 256 & 28 \\
PARBOIL & mri-q & ComputeQ\_GPU & 256 & 24 \\
PARBOIL & sgemm & mysgemmNT     & 128 &48 \\
PARBOIL & stencil & block2D\_hybrid\_coarsen\_x & 512   & 28 \\  
\hline
\end{tabular}
\label{table:set1}
\vskip -5mm
\end{table}

\begin{table}[t]
\caption{Set-2: Benchmarks that are limited by scratchpad memory}
\vskip -1mm
\renewcommand{\arraystretch}{.8}
\begin{tabular}{@{}l@{\ }l@{\ }l@{\ }l@{\ }l@{}}
\hline\hline
Benchmark &  Application & Kernel & Block & Scratch-\\
          &              &        & Size  & pad Size\\
\hline
CUDA-SDK  & convolutionSeparable & convolutionRows- & 64 & 2560 \\
& (CONV1) & Kernel& &  \\
CUDA-SDK  & convolutionSeparable & convolutionColumns- & 128 & 5184 \\
& (CONV2) & Kernel& &  \\
RODINIA & lavamd & kernel\_gpu\_cuda & 128 & 7200 \\
RODINIA & nw (NW1) & needle\_cuda\_shared\_1 & 16& 2180 \\
RODINIA & nw (NW2) & needle\_cuda\_shared\_2 & 16 & 2180 \\
RODINIA & srad\_v2 (SRAD1) & srad\_cuda\_1 & 256 & 6144 \\
RODINIA & srad\_v2 (SRAD2) & srad\_cuda\_2 & 256        & 5120 \\  
\hline
\end{tabular}
\label{table:set2}
\vskip -2mm
\end{table}

\begin{table}[t]
\caption{Set-3: Benchmark that are limited by threads or blocks}
\vskip -1mm
\renewcommand{\arraystretch}{.8}
\begin{tabular}{l@{\ }l@{\ }l@{\ }l@{\ }l@{\ }}
\hline\hline
Benchmark &  Application & Kernel &  Limited by\\
\hline
RODINIA  & backprop & bpnn\_layerforward\_CUDA & Threads \\
GPGPU-Sim& BFS      & Kernel                   & Threads \\
RODINIA  & gaussian & FAN2                     & Blocks \\
GPGPU-Sim&      NN  & executeSecondLayer       & Blocks \\
\hline
\end{tabular}
\label{table:set3}
\vskip -7mm
\end{table}

\subsection{Experimental Results} \label{sec:experiments}
We measured the performance of our approach using the number of Instructions executed Per Cycle (IPC), number of stall cycles, and number of idle cycles and compared it with the baseline GPGPU-Sim~\cite{GPGPU-Sim} implementation.

\begin{figure*}[t!]
\begin{center}
  \renewcommand{\arraystretch}{.5}
  \begin{tabular}{@{}c@{}c@{}c@{}c}
    {\small \ \ \ \ \ (a)} \hspace*{-10mm}\includegraphics[scale=0.5]{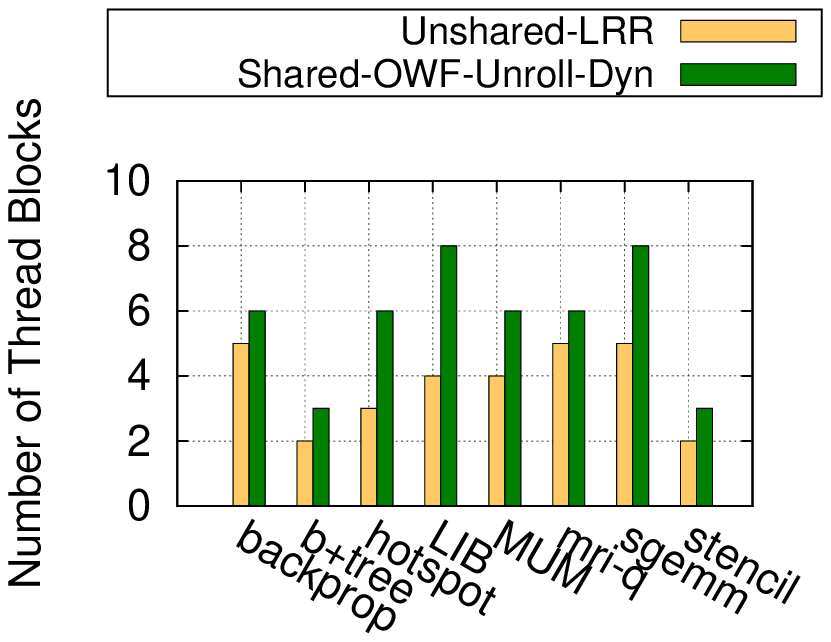} &
    {\small \ \ \ \ \ (b)} \hspace*{-10mm}\includegraphics[scale=0.5]{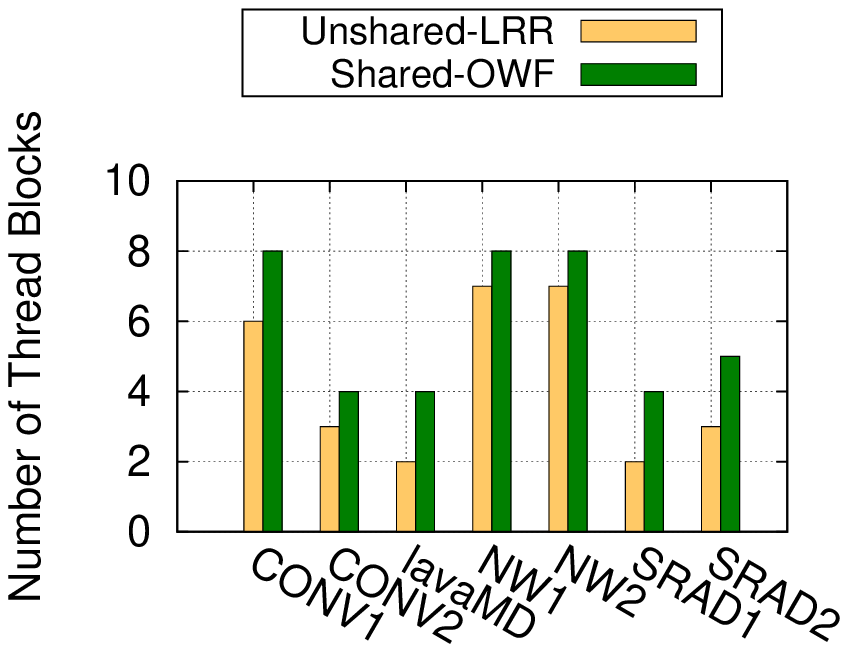} &
    {\small \ \ \ \ \ (c)} \hspace*{-10mm}\includegraphics[scale=0.5]{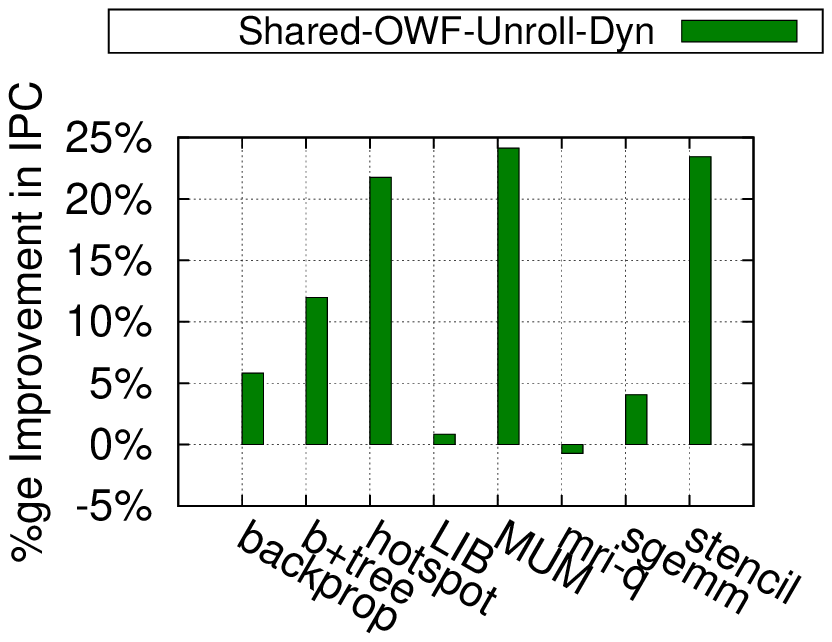} &		        
    {\small \ \ \ \ \ (d)} \hspace*{-10mm}\includegraphics[scale=0.5]{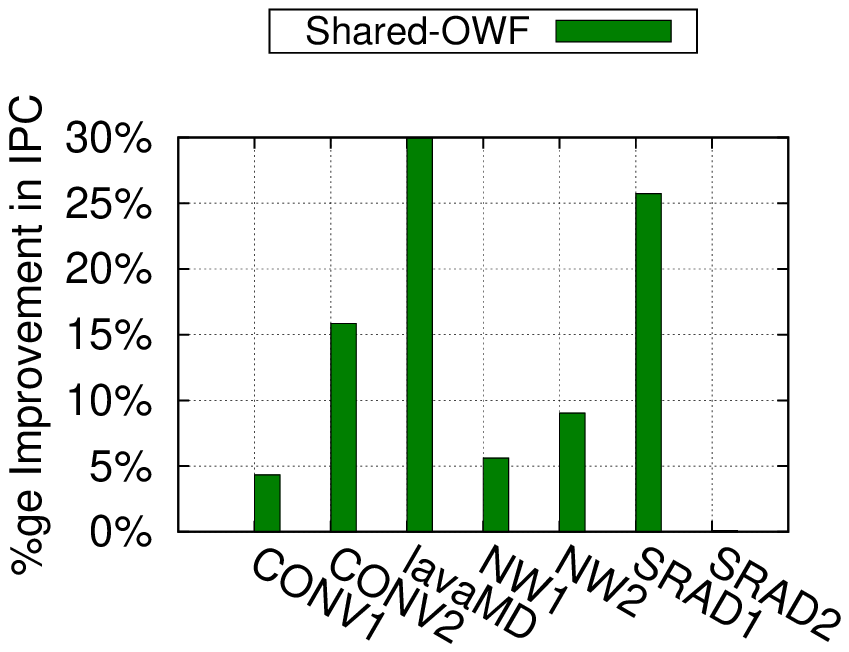} 
  \end{tabular} \\
  \caption{ Comparing the number of resident thread blocks with baseline implementation (a) Registers (c) Scratchpad Memory.\\
    Performance improvement in IPC when compared with baseline implementation (b) for Registers (d) for Scratchpad Memory \label{fig:LRRvsOWF-DYN}}        
\end{center} 
\vskip -2mm
\end{figure*}

We first show that resource sharing helps in increasing the number of thread blocks launched for the applications in Set-1 and Set-2. In Figure~\ref{fig:LRRvsOWF-DYN}(a), we compare the effective number of thread blocks launched by register sharing approach (denoted as Shared-OWF-Unroll-Dyn) with that of baseline implementation (denoted as Unshared-LRR). For applications \emph{MUM}, \emph{backprop}, \emph{hotspot}, and \emph{mri-q} our approach is able to launch 6 thread blocks (i.e, 1536 threads), which is the maximum limit on the number of resident threads per SM. Applications \emph{stencil} and \emph{b+tree} launch 3 thread blocks per SM, compared to 2 in the baseline approach. For applications \emph{LIB} and \emph{sgemm} our approach is able to launch 8 thread blocks per SM, which is the maximum limit on the number of resident thread blocks. 

Similarly in Figure~\ref{fig:LRRvsOWF-DYN}(b), we compare the number of resident thread blocks launched by scratchpad sharing (labeled as Shared-OWF) with baseline approach. \emph{CONV1}, \emph{NW1}, and \emph{NW2} launch upto 8 thread blocks per SM, which is the maximum limit on the number of resident thread blocks. 

\begin{figure*}[t]
\begin{center}
  \renewcommand{\arraystretch}{.5}
  \begin{tabular}{@{}c@{}c@{}c@{}c}
    {\small \ \ \ \ \ (a)} \hspace*{-10mm}\includegraphics[scale=0.5]{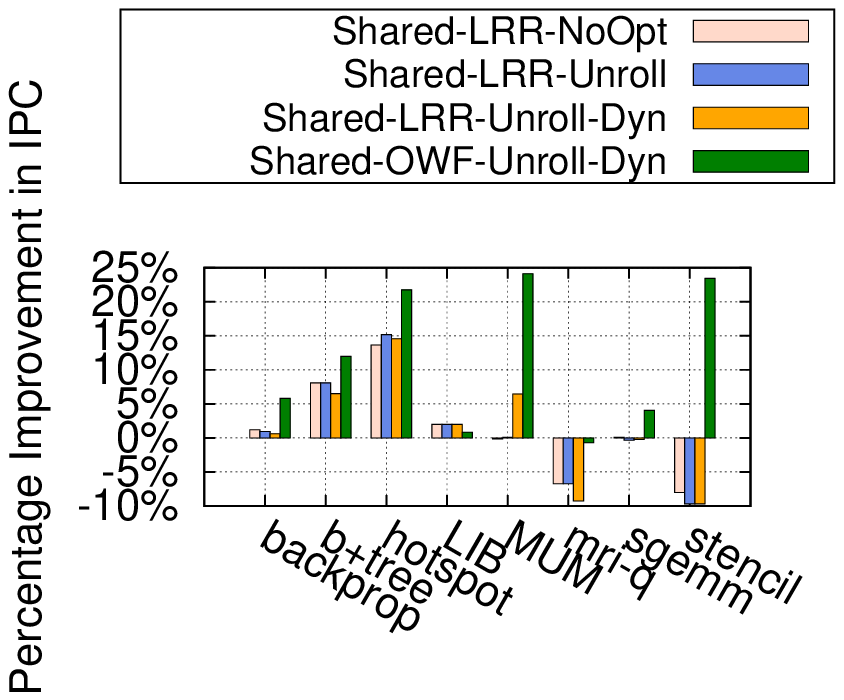} &        
    {\small \ \ \ \ \ (b)} \hspace*{-10mm}\includegraphics[scale=0.5]{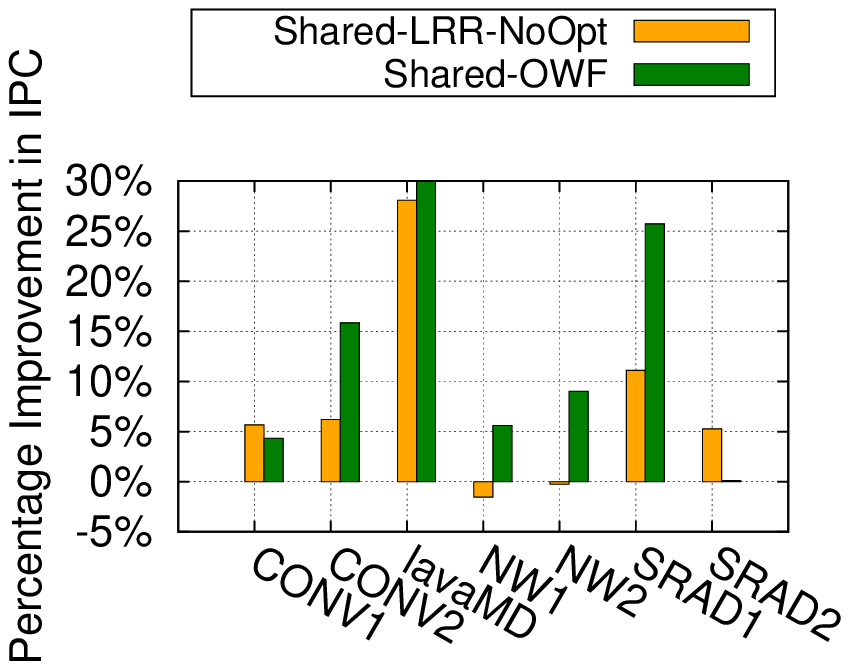} &
    {\small \ \ \ \ \ (c)} \hspace*{-10mm}\includegraphics[scale=0.5]{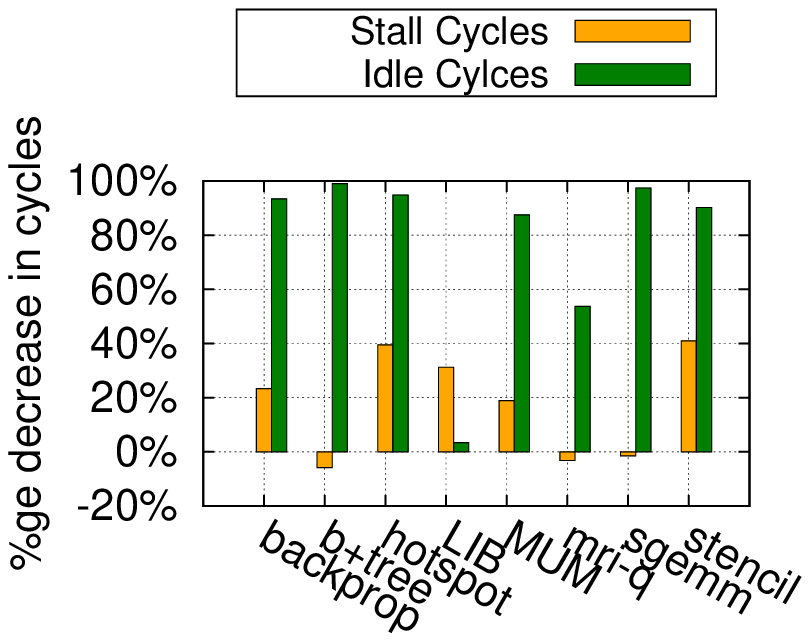} &
    {\small \ \ \ \ \ (d)} \hspace*{-10mm}\includegraphics[scale=0.5]{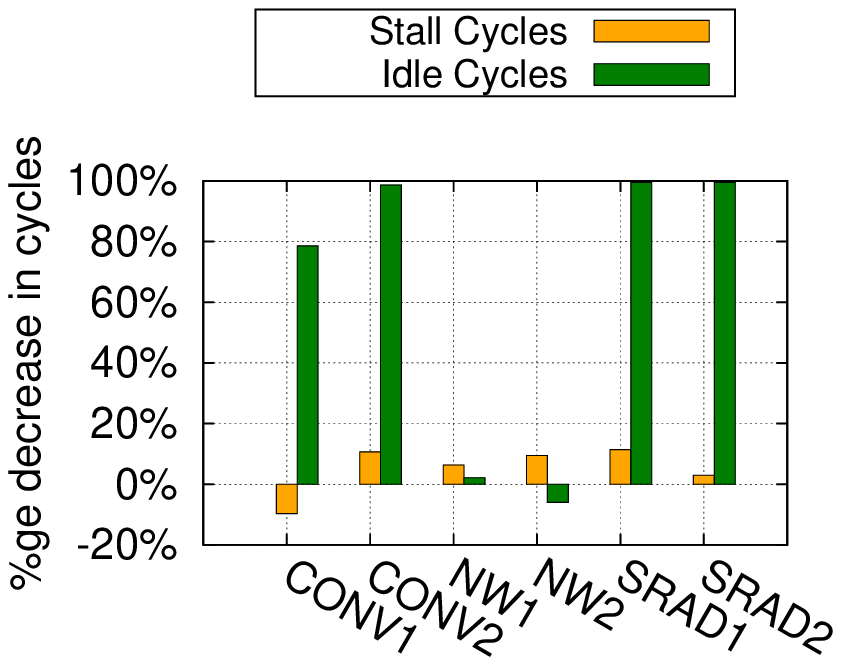} 
  \end{tabular}        
  \caption{ Performance analysis of optimizations for (a) Register Sharing (c) Scratchpad  sharing percentage decrease in stalls and idle cycles for (a) Register Sharing (c) Scratchpad  sharing \label{fig:OptimizationsAndStalls}. Note that lavaMD is not shown in (c) as it has zero stall cycles in baseline approach and 259 cycles in shared-OWF approach. It shows 49.5\% decrease in idle cycles.}
\end{center}    
\vskip -2mm
\end{figure*}

\begin{figure*}[t]
\begin{center}
  \renewcommand{\arraystretch}{0.5}
  \begin{tabular}{@{}c@{}c@{}c@{}c}
    {\small \ \ \ \ \ (a)} \hspace*{-10mm}\includegraphics[scale=0.5]{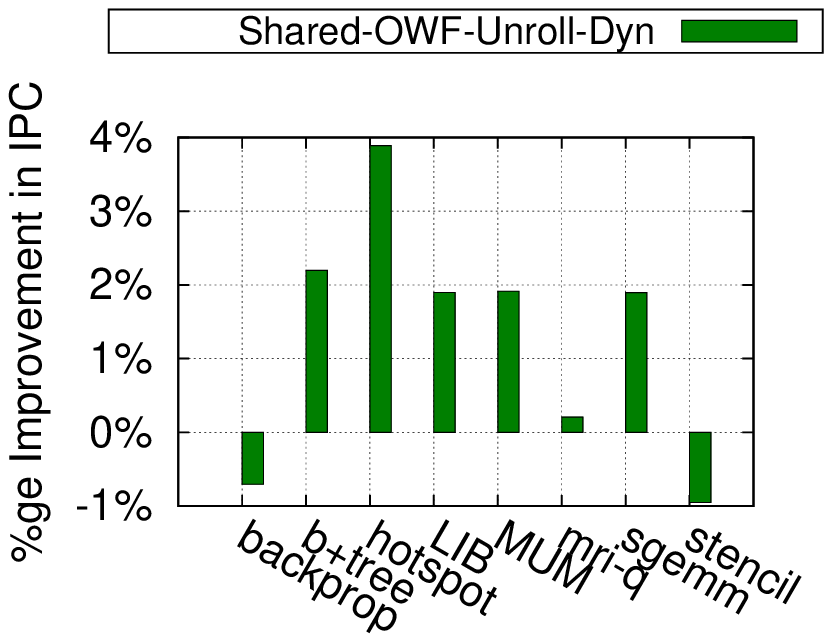} &
    {\small \ \ \ \ \ (b)} \hspace*{-10mm}\includegraphics[scale=0.5]{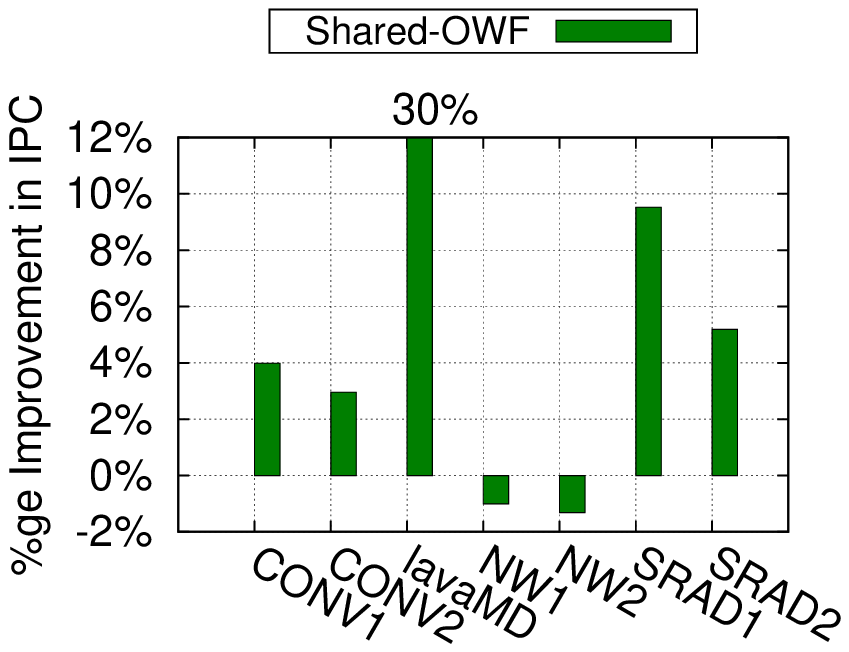} &
    {\small \ \ \ \ \ (c)} \hspace*{-10mm}\includegraphics[scale=0.5]{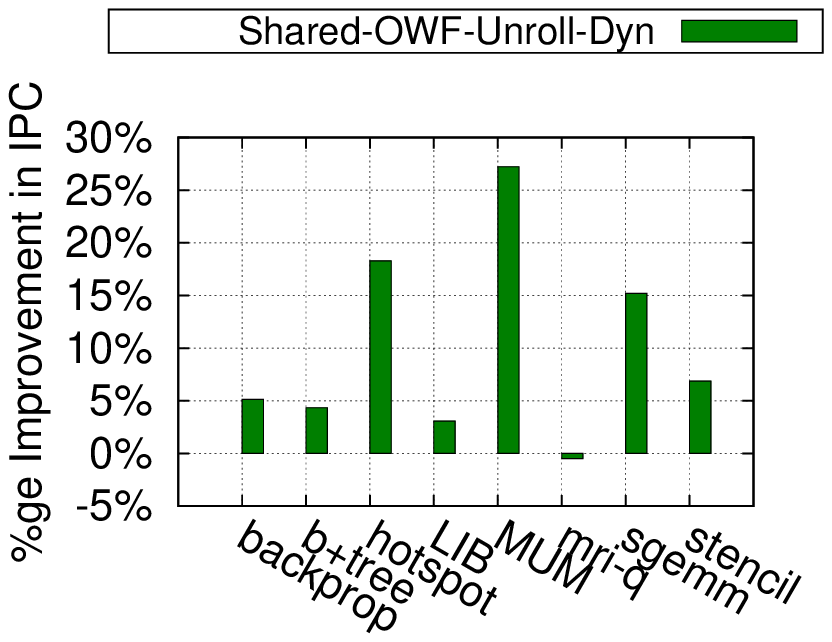} &
    {\small \ \ \ \ \ (d)} \hspace*{-10mm}\includegraphics[scale=0.5]{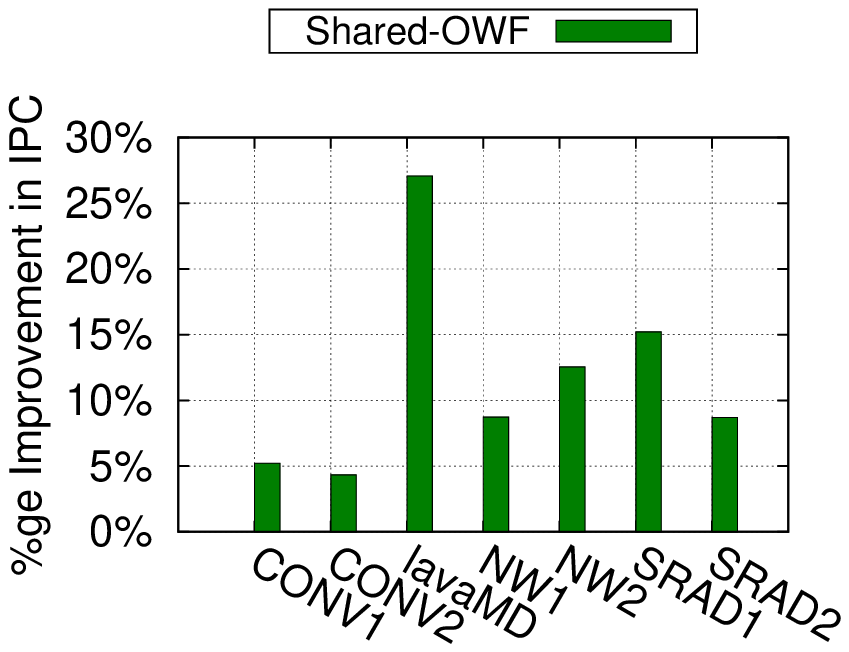}
  \end{tabular}        
  \caption{Performance comparison of (a) Scratchpad  sharing (b) Register sharing with GTO (baseline) scheduler \\
    Performance comparison of (a) Scratchpad  sharing (c) Register sharing with 2-Level (baseline) scheduler\label{fig:GTO2LevelvsOWF-DYN}}
\end{center} 
\vskip -7mm
\end{figure*}

Figure~\ref{fig:LRRvsOWF-DYN}(c) shows the improvement in IPC with register sharing over the baseline LRR (Loose Round Robin) implementation. We observe that applications \emph{b+tree}, \emph{hotspot}, \emph{MUM}, and \emph{stencil} achieve significant speedups of 11.98\%, 21.76\%, 24.14\%, and 23.45\% respectively. Similarly Figure~\ref{fig:LRRvsOWF-DYN}(d) shows the performance improvement in IPC with scratchpad sharing. \emph{CONV1}, \emph{lavaMD}, and \emph{SRAD2} achieve significant speedups of 15.85\%, 29.96\%, and 25.73\% respectively. These applications leverage all our optimizations to perform better.  The performance improvement in IPC for \emph{lavaMD} is due to two reasons (1) The number of resident thread blocks launched by our approach is twice that of baseline implementation (2) No instruction that uses scratchpad memory location falls into shared scratchpad, hence all the additional thread blocks execute instructions without waiting for shared thread blocks. Though \emph{LIB} launches 8 thread blocks per SM when register sharing is enabled, it improves only by 0.84\%. It is due to increase in L2 cache misses that are caused by additional shared blocks. The benchmarks \emph{backprop} and \emph{sgemm} achieve modest improvements of 5.82\% and 4.06\% respectively with register sharing. Similarly, \emph{CONV1}, \emph{NW1}, and \emph{NW2} show  improvements of 4.33\%, 5.62\%, and 9.03\% respectively with scratchpad sharing. \emph{mri-q} slows down by 0.72\%, because additional shared blocks increase L1 cache misses and hence increase the number of stalls. \emph{SRAD2} shows improvement only upto 0.1\%, because it has a barrier instruction (placed next to an instruction that accesses shared scratchpad memory) which limits the progress of shared threads that do not access any shared scratchpad location.

In Figure~\ref{fig:OptimizationsAndStalls}(a), we show the effectiveness of our proposed optimizations for register sharing by comparing them with the baseline approach. First we compare the results of register sharing approach when we do not use any optimization and use the existing baseline LRR scheduling policy (labeled Shared-LRR-NoOpt). Consider the application  \emph{hotspot}, it achieves a speedup of 13.65\% even without using any optimization because the additional thread blocks launched by our approach help in hiding execution latencies. With register unrolling optimization (labeled Shared-LRR-Unrolled), we further see an improvement up to 15.18\%, because register unrolling helps to increase the usage of unshared registers before they start using shared registers. Hence the application can execute more instructions before it accesses shared registers. When we enable the dynamic warp execution (labeled Shared-LRR-Unrolled-Dyn), we see an improvement only up to 14.58\%, because it limits the execution of memory instructions from non-owner warps. However when we apply the OWF optimization (labeled Shared-OWF-Unrolled-Dyn), the application speeds further up to 21.76\%. In the presence of OWF optimization, the priority of non-owner warps decreases compared to the other warps. Hence the memory instructions issued by non-owner warps do not interfere with the other warps, which minimizes the L1/L2 cache misses. We can see that \emph{b+tree} also behaves similar to \emph{hotspot} application in terms of performance gain by varying the optimizations.

For application \emph{MUM}, when we do not use any optimization, there is a slow down of 0.15\%. We observe that, increase in the resident thread blocks leads to increase in the L1 and L2 cache misses that arise by issuing memory instructions from the non-owner warps. Though we see an increase in the L1/L2 cache misses, the other instructions issued by the non-owner warps help in minimizing the stall cycles. With register unrolling optimization, we see a slight improvement (0.08\%). When we apply the dynamic warp execution, it shows a speed up of 6.45\%. From this we analyze that dynamic warp execution reduces the additional stall cycles produced by issuing memory instructions from the non-owner warps. Further with OWF optimization, performance improvement goes up to 24.14\%, because of decrease in interference from the non-owner warps.

\emph{LIB} shows an improvement of 2\% in the presence of sharing with no optimizations. We observe the same performance even with unrolling optimization, because the number of instructions that use unshared registers before the first instruction using shared registers is exactly the same as without the optimization. With dynamic warp execution, we still observe the same results, since in this application all the the owner warps have completed executing all instructions before any non-owner warp starts issuing any memory instructions. With OWF optimization, we observe a small degradation because of increase in the number of stall cycles compared to the LRR scheduling policy.

The benchmarks \emph{sgemm}, \emph{backprop}, and \emph{stencil} achieve good improvements only when OWF optimization is enabled. Since instructions issued by non-owner warps execute with the least priority, they  do not interfere with other warps and hence minimize L1/L2 cache misses. We do not see any performance improvement with \emph{mri-q}, because the additional thread blocks increase L1 cache misses with our approach. However the slow down was reduced to 0.72\% in the presence of all the optimizations.

To summarize, memory bound applications, like MUM, take advantage of our proposed sharing approach in the presence of dynamic warp execution and OWF optimizations. Whereas, compute bound applications, like hotspot, perform better even without any optimizations, and they further improve with the OWF optimization. 

In Figure~\ref{fig:OptimizationsAndStalls}(b), we show the effect of OWF optimization on scratchpad sharing. \emph{lavaMD} shows an improvement of 28\% even without any optimization (labeled shared-LRR-NoOpt). It is because the additional thread blocks do not access any memory location which belong to shared scratchpad memory. \emph{CONV1}, \emph{CONV2}, \emph{SRAD1}, and \emph{SRAD2} applications show improvements of 5.68\%, 6.21\%, 11.1\%, and 5.28\% respectively without applying optimization, which is due to additional thread blocks that help in hiding the latencies.

With OWF optimization, \emph{CONV2}, \emph{NW1}, \emph{NW2}, and \emph{SRAD2} applications improve upto 15.85\%, 5.62\%, 9.03\%, and 25.73\% respectively. Since OWF optimization schedules the owner warps efficiently, it helps in minimizing stall cycles thus improving IPC value. \emph{lavaMD} improves upto 30\%, since it has more benefit with sharing than OWF optimization. \emph{CONV1} and \emph{SRAD2} perform better when no optimization is applied, because these applications observe more number of L1, L2 misses and stall cycles with OWF optimization when compared to no optimization.

In Figure~\ref{fig:OptimizationsAndStalls}(c) and (d), we report percentage decrease in the number of idle cycles (Cycle in which all the available warps are issued, but no warp is ready to execute) and stall cycles (Pipeline stall) compared to the baseline implementation. We observe that, all applications show reduction in the number of idle cycles (max up to 99\%). This is expected because with increase in the number of thread blocks, the number of instructions that are ready to execute also increases. For applications \emph{MUM}, \emph{LIB}, \emph{backprop}, \emph{hotspot}, and \emph{stencil} the number stall cycles also reduce with our register approach.  Similarly for \emph{CONV2}, \emph{NW1}, \emph{NW2}, \emph{SRAD1}, and \emph{SRAD2} applications the number of stall cycles reduce with scratchpad sharing. It indicates the additional thread blocks launched with our approach hide the long execution latencies in a better way. We observe an increase in the stall cycles for applications \emph{b+tree} and \emph{stencil}. However, since the number of idle cycles have significantly reduced, overall we see a benefit with our approach. For \emph{mri-q}, the number of stall cycles increase with our approach due to increase in the number of L1 cache misses. \emph{lavaMD} shows increase the number of stall cycles by 259, this is because the additional threads launched by our approach are waiting for execution units (SP units) to become ready. Also with \emph{CONV1}, we see an increase in number of stalls, since our approach has more L1 cache misses compared to the baseline approach.

In Figures~\ref{fig:GTO2LevelvsOWF-DYN}(a) and (b), we show the performance improvement in register and scratchpad sharing approach over GTO (Greedy Then Old) scheduler respectively. We observe that our approach shows an improvement up to 3.9\% with register sharing and shows an improvement upto 30\% with scratchpad sharing. Further, as shown in Figure~\ref{fig:GTO2LevelvsOWF-DYN}(c) and (d) we observe an improvement up to 27.22\% in IPC with register sharing, and improvement upto 27.08\% with scratchpad sharing over the two-level scheduling policy.

\begin{figure}[t]
\begin{center}
  \renewcommand{\arraystretch}{0.5}
  \begin{tabular}{@{}c@{}c}
    {\small \ \ \ \ \ (a)} \hspace*{-10mm}\includegraphics[scale=0.45]{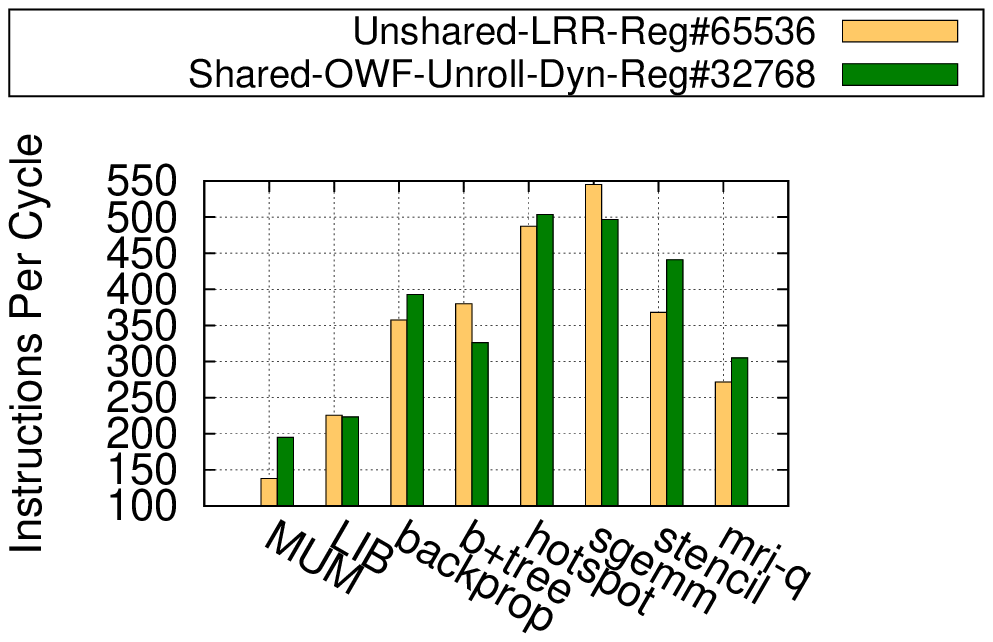} &
    {\small \ \ \ \ \ (b)} \hspace*{-10mm}\includegraphics[scale=0.45]{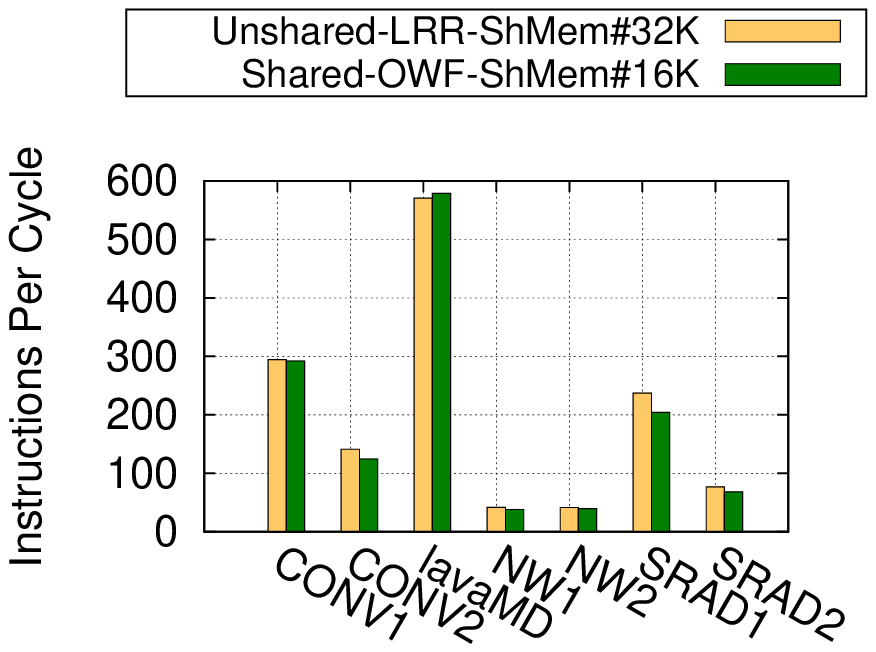} 
  \end{tabular}        
  \vskip -2mm
  \caption{Comparison with LRR that uses twice the number of (a) Registers (b) Scratchpad \label{fig:TwiceResourceAndSet3}}
  \vskip -9mm
\end{center} 
\vskip -4mm
\end{figure}

We also measure the effectiveness of resource sharing mechanism by comparing it with LRR Scheduler that uses twice the number of resources. In Figure~\ref{fig:TwiceResourceAndSet3}(a), the baseline approach (labeled Unshared-LRR-Reg\#65536) uses 64K registers, where as our approach uses only 32K registers. Even with an increase in the number of registers and hence an increase in the number of resident thread blocks in the baseline approach, our approach performs better in 5 out of 8 applications. For example, application
\emph{MUM} performs better with our approach, though we see the same number of thread blocks (6) in both the approaches as dynamic warp execution optimization helps minimizing the stalls produced by the additional thread blocks. Applications \emph{sgemm}, \emph{b+tree}, and \emph{LIB} perform better with the baseline approach due to an increase in the number of resident thread blocks and hence an increase in the number of active warps. In Figure~\ref{fig:TwiceResourceAndSet3}(b), we compare scratchpad sharing approach that uses 16K bytes of memory with that of baseline approach that uses 32K byes of memory. From the Figure we observe that, performance of \emph{CONV1}, \emph{NW1}, and \emph{NW2} is comparable to that of baseline approach, because our approach can launch the number of resident thread blocks which is equal to that of baseline approach. \emph{lavaMD} performs better than baseline approach, because of sharing helps in minimizing latencies, and also OWF optimization helps in scheduling the warps efficiently. \emph{CONV2}, \emph{SRAD1}, and \emph{SRAD2} degrades with our approach, because (1) number of resident thread blocks in our approach is less than that of baseline approach, and (2) the number of stall cycles observed in our approach is more than that of baseline implementation. 

\subsubsection{Effect of sharing on performance}
\begin{table}[t]
\caption{Effect on IPC with register sharing}
\vskip -1mm \centering
\renewcommand{\arraystretch}{.8}
\begin{tabular}{l@{} c c c c c c c@{}}
\hline\hline
\% Sharing\ \  &  0\% &  10\% &  30\% &  50\% &  70\% &  90\% \\
\hline
backprop & 389.9 &  389.9 &  389.9 &  389.9 &  394.1 &  392.8 \\
b+tree &  318.5 &  318.5 &  318.5 &  323.3 &  326.1 &  326.1 \\
hotspot &  489.5 &  489.5 &  489.5 &  475.2 &  476.9 &  503.59 \\
LIB &  218.0 &  218.0 &  203.0 &  203.0 &  216.3 &  223.3 \\
MUM &  190.5 &  190.5 &  190.5 &  192.1 &  192.4 &  194.9 \\
mri-q &  303.7 &  303.7 &  303.7 &  303.7 &  305.3 &  305.0 \\
sgemm &  490.6 &  490.6 &  490.6 &  490.6 &  446.3 &  496.7 \\
stencil &  448.2 &  448.2 &  448.2 &  448.2 &  448.2 &  440.8 \\ [1ex]
\hline
\end{tabular}
\label{table:threshold}
\vskip -2mm
\end{table}

\begin{table}[t]
\caption{Effect on resident thread blocks with register sharing}
\vskip -1mm \centering
\renewcommand{\arraystretch}{.8}
\begin{tabular}{l c c c c c c c}
\hline\hline
\% Sharing\ \  &  0\% &  10\% &  30\% &  50\% &  70\% &  90\% \\
\hline
backprop & 5 &  5 &  5 &  5 &  6 &  6 \\
b+tree &  2 &  2 &  2 &  3 &  3 &  3 \\
hotspot &  3 &  3 &  3 &  4 &  4 &  6 \\
LIB &  4 &  4 &  5 &  5 &  6 &  8 \\
MUM &  4 &  4 &  4 &  5 &  5 &  6 \\
mri-q &  5 &  5 &  5 &  5 &  6 &  6 \\
sgemm &  5 &  5 &  5 &  5 &  6 &  8 \\
stencil &  2 &  2 &  2 &  2 &  2 &  3 \\ [1ex]
\hline
\end{tabular}
\label{table:threshold_blocks}
\vskip -5mm
\end{table}

\begin{table}[t]
\caption{Effect on IPC with scratchpad sharing}
\vskip -1mm \centering
\renewcommand{\arraystretch}{.8}
\begin{tabular}{l@{} c c c c c c c@{}}
\hline\hline
\% Sharing\ \  &  0\% &  10\% &  30\% &  50\% &  70\% &  90\% \\
\hline
CONV1 & 280.33 & 280.33 & 280.33 & 280.33 & 288.82 & 292.24 \\
CONV2 & 119.29 & 119.29 & 119.29 & 119.29 & 119.02 & 124.6 \\
lavaMD & 452.29 & 452.29 & 452.29 & 452.29 & 452.29 & 578.85 \\
NW1 & 39.96 & 39.96 & 39.96 & 38.67 & 38.37 & 38.37 \\
NW2 & 41.93 & 41.93 & 41.93 & 42.14 & 40.54 & 39.72 \\
SRAD1 & 188.13 & 188.13 & 188.13 & 229.38 & 208.27 & 204.32 \\
SRAD2 & 63.48 & 63.48 & 63.48 & 63.52 & 63.62 & 68.29 \\ [1ex]
\hline
\end{tabular}
\label{table:shared_threshold}
\vskip -2mm
\end{table}

\begin{table}[t]
\caption{Effect on resident thread blocks with scratchpad sharing}
\vskip -1mm \centering
\renewcommand{\arraystretch}{.8}
\begin{tabular}{l c c c c c c c}
\hline\hline
\% Sharing\ \  &  0\% &  10\% &  30\% &  50\% &  70\% &  90\% \\
\hline
CONV1 & 6 & 6 & 6 & 6 & 7 & 8 \\
CONV2 & 3 & 3 & 3 & 3 & 3 & 4 \\
lavaMD & 2 & 2 & 2 & 2 & 2 & 4 \\
NW1 & 7 & 7 & 7 & 8 & 8 & 8 \\
NW2 & 7 & 7 & 7 & 8 & 8 & 8 \\
SRAD1 & 2 & 2 & 2 & 3 & 4 & 4 \\
SRAD2 & 3 & 3 & 3 & 3 & 3 & 5 \\ [1ex]
\hline
\end{tabular}
\label{table:shared_threshold_blocks}
\vskip -5mm
\end{table}

In Table~\ref{table:threshold} and Table~\ref{table:shared_threshold} , we analyze the performance of resource sharing  approach with the amount of sharing. From the results we observe that, most of the applications perform better when the amount of sharing is 90\%. It is because, as shown in Tables~\ref{table:threshold_blocks} and ~\ref{table:shared_threshold_blocks} , with increase in the amount of resource sharing, the number of resident thread blocks will increase. These resident thread blocks help in hiding long latencies and hence help in achieving high throughput. From the Tables~\ref{table:threshold} and ~\ref{table:shared_threshold}, we also notice that, all applications behave same at 0\% and 10\% sharing. At these percentages of sharing, the number of resident thread blocks with our approach is same as that of baseline implementation. Hence at run time, our approach decides to launch all the thread blocks in the unsharing mode. Since all these blocks are in unsharing mode, all the warps become unshared warps. In this case, OWF optimization uses dynamic warp ids to schedule warps and achieves higher performance than the baseline approach. \emph{SRAD2} (Table ~\ref{table:shared_threshold}) performs better at 50\%, because at this sharing, the number of instructions that get executed before they start enter the shared scratchpad memory region is more than that at 90\%. Also  at 50\% sharing, these extra instructions belong to loop statements, hence we observe more IPC values.

\subsubsection{Performance analysis of Set-3 benchmarks}

\begin{figure}[t]
\begin{center}
  \renewcommand{\arraystretch}{0.5}
  \begin{tabular}{@{}c@{}c}
    \includegraphics[scale=0.5]{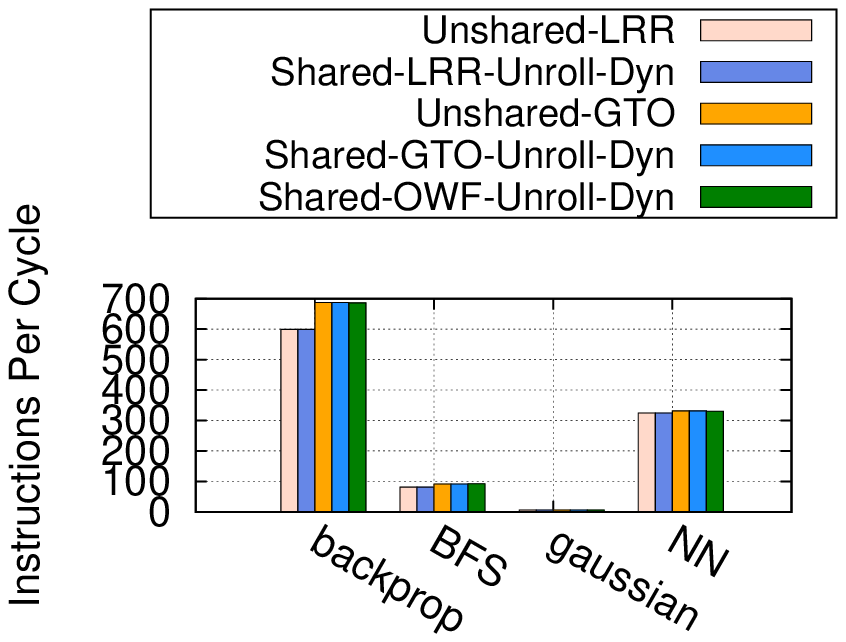} &
    \includegraphics[scale=0.5]{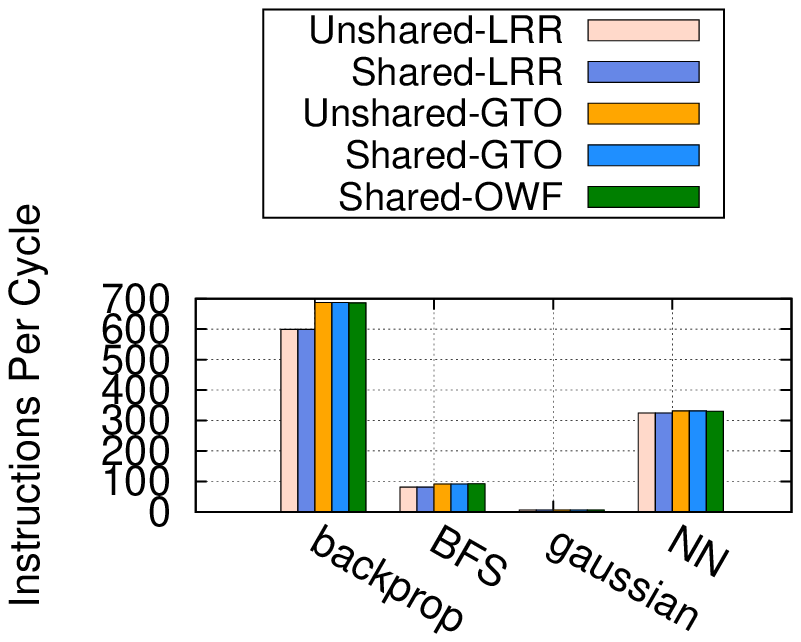} \\
    \small{(a)} & \small{(b)}
  \end{tabular}        
  \vskip -1mm
  \caption{ Performance analysis of Set-3 applications for (a) Register sharing (b) Scratchpad  sharing \label{fig:TwiceResourceAndSet3}}
\end{center} 
\vskip -5mm
\end{figure}

The performance of register sharing and scratchpad sharing approach for the Set-3 applications
(Table~\ref{table:set3}) is presented in Figures~\ref{fig:TwiceResourceAndSet3}(c) and (d) respectively. As discussed earlier, these applications are
not limited by the number of available resources but due to other factors
such as the number of threads, thread blocks, etc.
We measure their performance when our approach uses (1) LRR scheduling policy, (2) GTO scheduling policy, and (3) OWF scheduling policy\footnote{%
We do not use two-level scheduling policy, because it cannot be directly integrated with our sharing approach %
}. From Figures~\ref{fig:TwiceResourceAndSet3}(c) and (d), we observe that our proposed resource sharing approach when used with LRR scheduling (labeled as Shared-LRR-Unroll-Dyn) performs exactly same as the baseline LRR scheduling (Unshared-LRR).
Since the number of thread blocks launched by the
applications are not limited by the resources, our approach
does not launch any additional thread blocks, and all the
thread blocks are in unsharing mode. Hence, it behaves
exactly similar to the baseline approach. Similarly, 
our approach when used with the GTO scheduling policy (Shared-GTO-Unroll-Dyn), 
performs exactly same as the
baseline approach that uses GTO scheduling policy without sharing (Unshared-GTO). Finally,
we observe that 
with OWF scheduling policy (Shown as Shared-OWF-Unroll-Dyn),
our approach is comparable to that of
Unshared-GTO implementation. In OWF optimization, the warps
are arranged according to priority of owner, unshared, and
non-owner warps. Since in this case, we don't launch any
additional thread blocks, all the thread blocks are in
unshared mode. Hence all the unshared warps are sorted
according to their dynamic warp id. So the performance of
Shared-OWF-Unroll-Dyn is similar to that of Unshared-GTO
implementation.

\section{Related Work} \label{sec:relatedwork}

Xiang et. al.~\cite{WarpLevelDivergence} discussed thread block level resource management. \cmt{They classified the resource under utilization problems as temporal and spatial. Temporal under utilization is caused due to differences in run times of warps of a thread block, whereas, spatial under utilization is caused because of unavailability of enough resources for a complete thread block at the time of launching a kernel.} They proposed a hardware solution to launch a partial thread block when there are not enough resources to launch a full thread block. Unlike our approach, their solution can have only one partial thread block running.\cmt{, whereas, our solution can have multiple thread blocks in shared mode and hence provides more opportunities to hide long latencies. Also, their solution to handle warp-divergence is complementary to our approach and can be used as and when warps from unshared thread blocks finish.} A patented register management scheme in~\cite{ondemand}, uses the concept of virtual registers, which are more than the actual physical registers,  and hence can launch more thread blocks than allowed by the physical registers. This mechanism can be combined with our proposed solution. Yang et. al.~\cite{SharedMemMultiplexing} propose hardware and software solutions to the problem caused by allocation and deallocation of shared memory
at the thread block granularity.

Warped Register File~\cite{WarpedRegFile} describes a solution to reduce the power consumption in register file by turning off unallocated registers. Gebhart et. al.~\cite{Unification} proposed a unified memory for register, scratchpad, and primary cache, which partitions resources of SM as per the application need. It requires lot of hardware changes to access unified storage, in contrast our approach requires fewer modifications to the hardware. Other works~\cite{Brunie,ThreadFrontiers,TBCompaction,Han,Meng,DualPathExec,CAPRI,Linearization} propose hardware and software solutions to improve throughput of GPUs by handling branch and thread divergence, but these are orthogonal to our approach.

Other techniques to improve GPU performance include reducing cache contention, improving DRAM bandwidth, hide long latencies, reduce energy consumption, etc. Rogers et. al.~\cite{Rogers} propose a cache conscious wave front scheduling algorithm which makes use of intra-wave front locality detector, focusing on shared L1 cache. A Two level warp scheduler~\cite{TwoLevel} proposed by Narasiman et. al. divides warps into groups and schedules the warps in each group in round robin manner to hide long  latencies in a better way. Gebhart et. al.~\cite{Gebhart} proposed energy efficient hierarchical register file storage and two level warp scheduler for high throughput processors. OWL~\cite{OWL} proposes various techniques to improve cache contention and DRAM bank level parallelism.

\section{Conclusions and Future Work} \label{sec:conclusion}
In this paper we proposed a technique that shares resources of SM  to effectively utilize the wasted resources by launching additional thread blocks in each SM. \cmt{These thread blocks help in hiding the long instruction execution latencies, hence improving the throughput of applications.} For effective utilization of these additional thread blocks, we proposed optimizations which further help in reducing the stalls produced in the system. \cmt{We implemented our approach for optimizing two GPU resources (1) Scratchpad memory (2) Registers.} We validated our approach for register sharing and for scratchpad sharing on several applications, and showed improvements \cmt{that the applications improve}up to maximum 24\% and average 11\% with register sharing, and maximum 30\% and average 12.5\% with scratchpad sharing.

In future, we plan to incorporate traditional compiler analysis and optimizations into our approach. For example,  live range analysis along with instruction reordering can be used to detect and release registers that are not used beyond a point. Such registers, if shared, can be used by the warp in the other thread block waiting for shared registers. We also plan to study the effect of various cache replacement policies on register sharing and use it to improve the throughput of memory bound applications. 


\renewcommand{\IEEEbibitemsep}{0.8pt}
\small

\bibliographystyle{IEEEtran}
\bibliography{RegSharing}
\end{document}